\documentclass[
aps,prc,twocolumn,
showpacs,preprintnumbers,
amsmath,amssymb,
floatfix,
superscriptaddress,
]{revtex4-1}

\usepackage{graphicx}           
\graphicspath{{./figures/}}     
\usepackage{dcolumn}            
\usepackage{multirow}
\usepackage{tabulary}
\usepackage{booktabs}           
\usepackage{bm}                 
\usepackage{xcolor}              
\usepackage{amsmath}
\usepackage{textcomp,gensymb}   
\usepackage[normalem]{ulem}     
\usepackage{xargs}              
\usepackage{aas_macros}         
 \usepackage{comment}

\usepackage[colorinlistoftodos]{todonotes}  
\newcommandx{\unsure}[2][1=]{\todo[linecolor=red,backgroundcolor=red!25,bordercolor=red,#1]{#2}}
\newcommandx{\change}[2][1=]{\todo[linecolor=blue,backgroundcolor=blue!25,bordercolor=blue,#1]{#2}}
\newcommandx{\info}[2][1=]{\todo[inline,linecolor=OliveGreen,backgroundcolor=OliveGreen!25,bordercolor=OliveGreen,#1]{#2}}
\newcommandx{\improvement}[2][1=]{\todo[linecolor=Plum,backgroundcolor=Plum!25,bordercolor=Plum,#1]{#2}}

\newcommand{\ph}{\ensuremath{^{31}\mathrm{P}}}
\newcommand{\sit}{\ensuremath{^{30}\mathrm{Si}}}
\newcommand{\he}{\ensuremath{^{3}\mathrm{He}}}
\newcommand{\spg}{$^{30}$Si($p$,$\gamma$)$^{31}$P}
\newcommand{\shd}{$^{30}$Si($^{3}$He,$d$)$^{31}$P}
\newcommand{\ver}{Vernotte \emph{et~al} \cite{Vernotte1990}}
\newcommand{\reaction}[6]{\nuc{#1}{#2}(#3,#4)\/\nuc{#5}{#6}}
\newcommand{\nuc}[2]{\ensuremath{^{#1}}{#2}}

\newcommand{\Ercm}[1]{\ensuremath{E_{r}^{\text{c.m.}}~=~#1}~keV}
\newcommand{\Ex}[1]{\ensuremath{E_{x}~=~#1}~keV}

\let\jnl=\rm
\def\reff@jnl#1{{\jnl#1}}

\def\aj{\reff@jnl{AJ}}                   
\def\actaa{\reff@jnl{Acta Astron.}}      
\def\araa{\reff@jnl{ARA\&A}}             
\def\apj{\reff@jnl{ApJ}}                 
\def\apjl{\reff@jnl{ApJ}}                
\def\apjs{\reff@jnl{ApJS}}               
\def\ao{\reff@jnl{Appl.~Opt.}}           
\def\apss{\reff@jnl{Ap\&SS}}             
\def\aap{\reff@jnl{A\&A}}                
\def\aapr{\reff@jnl{A\&A~Rev.}}          
\def\aaps{\reff@jnl{A\&AS}}              
\def\azh{\reff@jnl{AZh}}                 
\def\baas{\reff@jnl{BAAS}}               
\def\bac{\reff@jnl{Bull. astr. Inst. Czechosl.}}
\def\caa{\reff@jnl{Chinese Astron. Astrophys.}}
\def\cjaa{\reff@jnl{Chinese J. Astron. Astrophys.}}
\def\icarus{\reff@jnl{Icarus}}           
\def\jcap{\reff@jnl{J. Cosmology Astropart. Phys.}}
\def\jrasc{\reff@jnl{JRASC}}             
\def\memras{\reff@jnl{MmRAS}}            
\def\mnras{\reff@jnl{MNRAS}}             
\def\na{\reff@jnl{New A}}                
\def\nar{\reff@jnl{New A Rev.}}          
\def\pra{\reff@jnl{Phys.~Rev.~A}}        
\def\prb{\reff@jnl{Phys.~Rev.~B}}        
\def\prc{\reff@jnl{Phys.~Rev.~C}}        
\def\prd{\reff@jnl{Phys.~Rev.~D}}        
\def\pre{\reff@jnl{Phys.~Rev.~E}}        
\def\prl{\reff@jnl{Phys.~Rev.~Lett.}}    
\def\pasa{\reff@jnl{PASA}}               
\def\pasp{\reff@jnl{PASP}}               
\def\pasj{\reff@jnl{PASJ}}               
\def\rmxaa{\reff@jnl{Rev. Mexicana Astron. Astrofis.}}%
\def\qjras{\reff@jnl{QJRAS}}             
\def\skytel{\reff@jnl{S\&T}}             
\def\solphys{\reff@jnl{Sol.~Phys.}}      
\def\sovast{\reff@jnl{Soviet~Ast.}}      
\def\ssr{\reff@jnl{Space~Sci.~Rev.}}     
\def\zap{\reff@jnl{ZAp}}                 
\def\nat{\reff@jnl{Nature}}              
\def\iaucirc{\reff@jnl{IAU~Circ.}}       
\def\aplett{\reff@jnl{Astrophys.~Lett.}} 
\def\apspr{\reff@jnl{Astrophys.~Space~Phys.~Res.}}
\def\bain{\reff@jnl{Bull.~Astron.~Inst.~Netherlands}} 
\def\fcp{\reff@jnl{Fund.~Cosmic~Phys.}}  
\def\gca{\reff@jnl{Geochim.~Cosmochim.~Acta}}   
\def\grl{\reff@jnl{Geophys.~Res.~Lett.}} 
\def\jcp{\reff@jnl{J.~Chem.~Phys.}}      
\def\jgr{\reff@jnl{J.~Geophys.~Res.}}    
\def\jqsrt{\reff@jnl{J.~Quant.~Spec.~Radiat.~Transf.}}
\def\memsai{\reff@jnl{Mem.~Soc.~Astron.~Italiana}}
\def\nphysa{\reff@jnl{Nucl.~Phys.~A}}   
\def\physrep{\reff@jnl{Phys.~Rep.}}   
\def\physscr{\reff@jnl{Phys.~Scr}}   
\def\planss{\reff@jnl{Planet.~Space~Sci.}}   
\def\procspie{\reff@jnl{Proc.~SPIE}}   

\begin{document}
 
\title{Experimental study of the \shd\ reaction and thermonuclear reaction rate of \spg}

\author{D.~S. Harrouz}
    \affiliation{Universit\'{e} Paris-Saclay, CNRS/IN2P3, IJCLab, 91405 Orsay, France}
\author{N. de S\'{e}r\'{e}ville}
    \email{nicolas.de-sereville@ijclab.in2p3.fr}
    \affiliation{Universit\'{e} Paris-Saclay, CNRS/IN2P3, IJCLab, 91405 Orsay, France}
\author{P. Adsley}
\altaffiliation[Present address:]{Cyclotron Institute and Department of Physics \& Astronomy, Texas A\&M University, College Station, Texas 77843, USA}
    \affiliation{School of Physics, University of the Witwatersrand, Johannesburg 2050, South Africa}
    \affiliation{iThemba Laboratory for Accelerator Based Sciences, Somerset West 7129, South Africa}
    
\author{F. Hammache}
    \affiliation{Universit\'{e} Paris-Saclay, CNRS/IN2P3, IJCLab, 91405 Orsay, France}
\author{R. Longland}
    \affiliation{North Carolina State University, Raleigh, NC 27695}
    \affiliation{Triangle Universities Nuclear Laboratory, Durham, NC 27708}
\author{B.~Bastin}
    \affiliation{Grand Acc\'el\'erateur National d'Ions Lourds (GANIL), CEA/DRF-CNRS/IN2P3, 
    Bd. Henri Becquerel, 14076 Caen, France}
\author{T. Faestermann}
    \affiliation{Physik Department E12, Technische Universit\"{a}t M\"{u}nchen, D-85748 Garching, Germany}
\author{R. Hertenberger}
    \affiliation{Fakult\"{a}t f\"{u}r Physik, Ludwig-Maximilians-Universit\"{a}t M\"{u}nchen, D-85748 Garching, Germany}
\author{M. La Cognata}
    \affiliation{Laboratori  Nazionali  del  Sud  -  Istituto  Nazionale  di  Fisica  Nucleare,  Via  Santa  Sofia  62,  95123  Catania,  Italy}
\author{L. Lamia}
    \affiliation{Laboratori  Nazionali  del  Sud  -  Istituto  Nazionale  di  Fisica  Nucleare,  Via  Santa  Sofia  62,  95123  Catania,  Italy}
    \affiliation{Dipartimento di Fisica e Astronomia E. Majorana, Univ. di Catania, Catania, Italy}
\author{A. Meyer}
    \affiliation{Universit\'{e} Paris-Saclay, CNRS/IN2P3, IJCLab, 91405 Orsay, France}
\author{S. Palmerini}
    \affiliation{Dipartimento  di  Fisica  e  Geologia,  Universit\`a  degli  Studi  di  Perugia,  Perugia,  Italy}
    \affiliation{Istituto  Nazionale  di  Fisica  Nucleare,  Sezione  di  Perugia,  Perugia,  Italy}
\author{R.~G. Pizzone}
    \affiliation{Laboratori  Nazionali  del  Sud  -  Istituto  Nazionale  di  Fisica  Nucleare,  Via  Santa  Sofia  62,  95123  Catania,  Italy}
\author{S. Romano}
    \affiliation{Laboratori  Nazionali  del  Sud  -  Istituto  Nazionale  di  Fisica  Nucleare,  Via  Santa  Sofia  62,  95123  Catania,  Italy}
    \affiliation{Dipartimento di Fisica e Astronomia E. Majorana, Univ. di Catania, Catania, Italy}
    \affiliation{Centro Siciliano di Fisica Nucleare e Struttura della Materia-CSFNSM, Catania, Italy}
\author{A. Tumino}
    \affiliation{Laboratori  Nazionali  del  Sud  -  Istituto  Nazionale  di  Fisica  Nucleare,  Via  Santa  Sofia  62,  95123  Catania,  Italy}
    \affiliation{Facolt\`a di Ingegneria e Architettura, Universit\`a degli Studi di Enna, Italy}
\author{H.-F. Wirth}
\affiliation{Fakult\"{a}t f\"{u}r Physik, Ludwig-Maximilians-Universit\"{a}t M\"{u}nchen, D-85748 Garching, Germany}
  
\date{\today}

\begin{abstract}
    \begin{description}
        \item[Background] Abundance anomalies in some globular clusters, such as the enhancement of potassium and the depletion of magnesium, can be explained in terms of an earlier generation of stars polluting the presently observed ones. It was shown that the potential range of temperatures and densities of the polluting sites depends on the strength of a few number of critical reaction rates. The \spg\ reaction has been identified as one of these important reactions.
        \item[Purpose] The key ingredient for evaluating the thermonuclear \spg\ reaction rate is the strength of the resonances which, at low energy, are proportional to their proton width. Therefore the goal of this work is to determine the proton widths of unbound $^{31}$P states.
        \item[Method] States in $^{31}$P were studied at the Maier-Leibnitz-Laboratorium using the one-proton \shd\ transfer reaction. Deuterons were detected with the Q3D magnetic spectrometer. Angular distribution and spectroscopic factors were extracted for 27 states, and proton widths and resonance strengths were calculated for the unbound states.
        \item[Results] Several \ph\ unbound states have been observed for the first time in a one-proton transfer reaction. Above 20~MK, the \spg\ reaction rate is now entirely estimated from the observed properties of \ph\ states. The reaction rate uncertainty from all resonances other than the \Ercm{149} resonance has been reduced down to less than a factor of two above that temperature. The unknown spin and parity of the \Ercm{149} resonance dominates the uncertainty in the rate in the relevant temperature range. 
        \item[Conclusion] The remaining source of uncertainty on the \spg\ reaction rate comes from the unknown spin and parity of the \Ercm{149} resonance which can change the reaction rate by a factor of ten in the temperature range of interest.
\end{description}

\end{abstract}

\maketitle

\section{Introduction}
Globular clusters are vital testing grounds for models of stellar evolution and early stages of the formation of galaxies. It is now well established that they consist of multiple star generations implying several episodes of star formation~\cite{Gratton2012, Piotto2007}. Spectroscopic observations show a typical anti-correlation of abundances between pairs of light elements such as C--N, O--Na and Mg--Al~\cite{2017A&A...601A.112P,*2010A&A...516A..55C,*2019AJ....158...14N,*2017A&A...608A..28P,*2009A&A...505..139C,*2012ApJ...761L..30V,*2012MNRAS.426.2889M,*2015ApJ...801...68M,*2015ApJ...810..148C,*2012ApJ...760...86C}. These abundance patterns have been detected in red giant stars where present day models show that the relatively low burning temperatures cannot modify the abundances of O, Na, Mg or Al~\cite{Powell1999}, indicating that these abundance anomalies come from an earlier stellar population which polluted the gas from which the currently observed globular cluster stars formed. So far, the nature of stellar polluters and the pollution process remains uncertain, and identifying them is a challenge to the understanding of how globular cluster and galaxy formation took place.

The NGC~2419 globular cluster deserves special attention since a Mg--K anticorrelation is also observed~\cite{Cohen2012} in addition to anticorrelation of lighter elements in giant stars. In this case the abundance of potassium is strongly enhanced while the magnesium one is depleted. This feature cannot be explained by low-temperature hydrogen burning ($T\approx75$~MK) in the polluter stars as was suggested to successfully explain both the O--Na and Mg--Al correlations in other globular clusters~\cite{Prantzos2007}. Indeed, the involved proton-capture reaction rates are too small because of the Coulomb barrier and the observed potassium overabundance cannot be reproduced. This indicates that a polluting star with hydrogen burning at elevated temperatures ($T\geq80$~MK) is therefore required in order to explain the Mg--K anticorrelation in NGC~2419. Recent studies relying on Monte Carlo network calculations have explored the temperature and density conditions of the polluter needed to reproduce these abundance anomalies~\cite{Iliadis2016,Dermigny2017}. These works suggest that the polluter material should be produced by hydrogen burning occurring in a temperature range of $120-200$~MK. However, this temperature range is sensitive to a few proton radiative capture reactions, \spg\ being one of them.

The \spg\ reaction rate is rather well constrained experimentally at high temperatures ($T \geq 1$~GK which are relevant for type Ia supernovae nucleosynthesis~\cite{Parikh2013}) where the strength of resonances above \Ercm{600} has been measured directly~\cite{Longland2010}. Hydrogen burning in the polluting stellar site in globular clusters takes place at lower temperatures  of $T=120-200$~MK where there are eight resonances in the \Ercm{100-500} range corresponding to these stellar temperatures. The strength of the \Ercm{482} resonance has been recently directly measured~\cite{Dermigny2020} resolving a discrepancy of a factor of two between previous measurements~\cite{Hoogenboom58,Riihonen1979}. Along with the reported value for the \Ercm{482} resonance, a direct strength measurement was reported for the \Ercm{422} resonance~\cite{Dermigny2020}, which is the lowest-energy resonance for which a direct measurement is available. Indirect determination of the proton widths provides an alternative approach to determining the reaction rates. For resonance energies corresponding to $T=120-200$~MK, the resonance strengths are directly proportional to the proton width since the proton widths are much smaller than the $\gamma$-ray widths~\cite{iliadisbook}. The proton width of the \Ercm{441} resonance has been determined indirectly from its proton spectroscopic factor, obtained from the one-proton \shd\ transfer reaction~\cite{Vernotte1990}. The proton widths of the other resonances were not determined due to the poor energy resolution not allowing to separate the \Ex{7719-7737} doublet, and the low statistics for the least populated states. The goal of this paper is to determine the proton widths of unbound \ph\ states within 500~keV above the $^{30}$Si+$p$ threshold at $S_p=7296.55(2)$~keV~\cite{nndc}.

In this work we report on a study of $^{31}$P states for excitation energies between 6.8 and 8.1 MeV. The \shd\ reaction was used to populate and identify the relevant states. From the analysis of the deuteron angular distributions the proton spectroscopic factors and the transferred angular momenta were obtained for each populated $^{31}$P level. Section \ref{sec:expt} describes the experimental setup and technique used, and Section \ref{sec:analysis} details the data analysis methodology. The results are presented in Section \ref{sec:results} and the \spg\ reaction rate is given in Section \ref{sec:rate}. We summarise and conclude in Section \ref{sec:conclusion}.

\section{Experiment}
\label{sec:expt}
The \shd\ reaction was studied at the Maier-Leibnitz-Laboratorium (MLL) in Garching, Germany. A $^{3}$He$^{2+}$ beam (I = 150 -- 200 enA) was accelerated to 25 MeV by the Tandem Van de Graaff accelerator, and transported to the target located at the object focal point of a Q3D magnetic spectrometer~\cite{DollingerQ3D}. The integrated charge was measured by a suppressed Faraday cup located at $0\degree$ downstream the target. An enriched (95\%) $^{30}$SiO$_2$ target of thickness 16(3)~$\micro$g/cm$^2$ on a 40(1)~$\micro$g/cm$^2$ carbon backing was used. In addition, natural silicon oxide and natural carbon targets were used to characterize the background, and to identify any contaminant reaction. The target thickness was determined using a Rutherford Back Scattering technique at the JANNuS/SCALP facility~\cite{Bacri2017} (Orsay, France) and at the Triangle Universities Nuclear Laboratory (Durham, North Carolina, USA).

The light reaction products entered the Q3D spectrometer through a smaller aperture ($\Delta\Omega=4.2$~msr) than the maximum one, in order to improve the energy resolution while maintaining a reasonable level of statistics. The reaction products were then momentum analysed and focused on its focal-plane detection system~\cite{Wirth2000}.
This system was equipped with a 0.9~m long detector, based on two single-wire proportional counters, one with a readout of 3~mm wide cathode strips for position ($\Delta x \sim 0.1$~mm) and energy-loss information, and a scintillator for the residual energy of the light particles. Particle identification was achieved by combining the aforementioned physical quantities and the deuterons were readily separated from other reaction products. 

The measurements were performed at the laboratory angles of $\theta_\mathrm{lab} = 6\degree, 10\degree, 12\degree, 16\degree, 20\degree, 23\degree$, and $32\degree$ with a precision of half an angular measurement gradient corresponding to $0.05\degree$. The magnetic field of the spectrometer was set at each detection angle to cover an excitation-energy range between $E_x=6.8$~MeV and 8.1~MeV.

\section{Data analysis}
\label{sec:analysis}
\begin{figure*}[!htpb]
    \includegraphics[width=\textwidth]{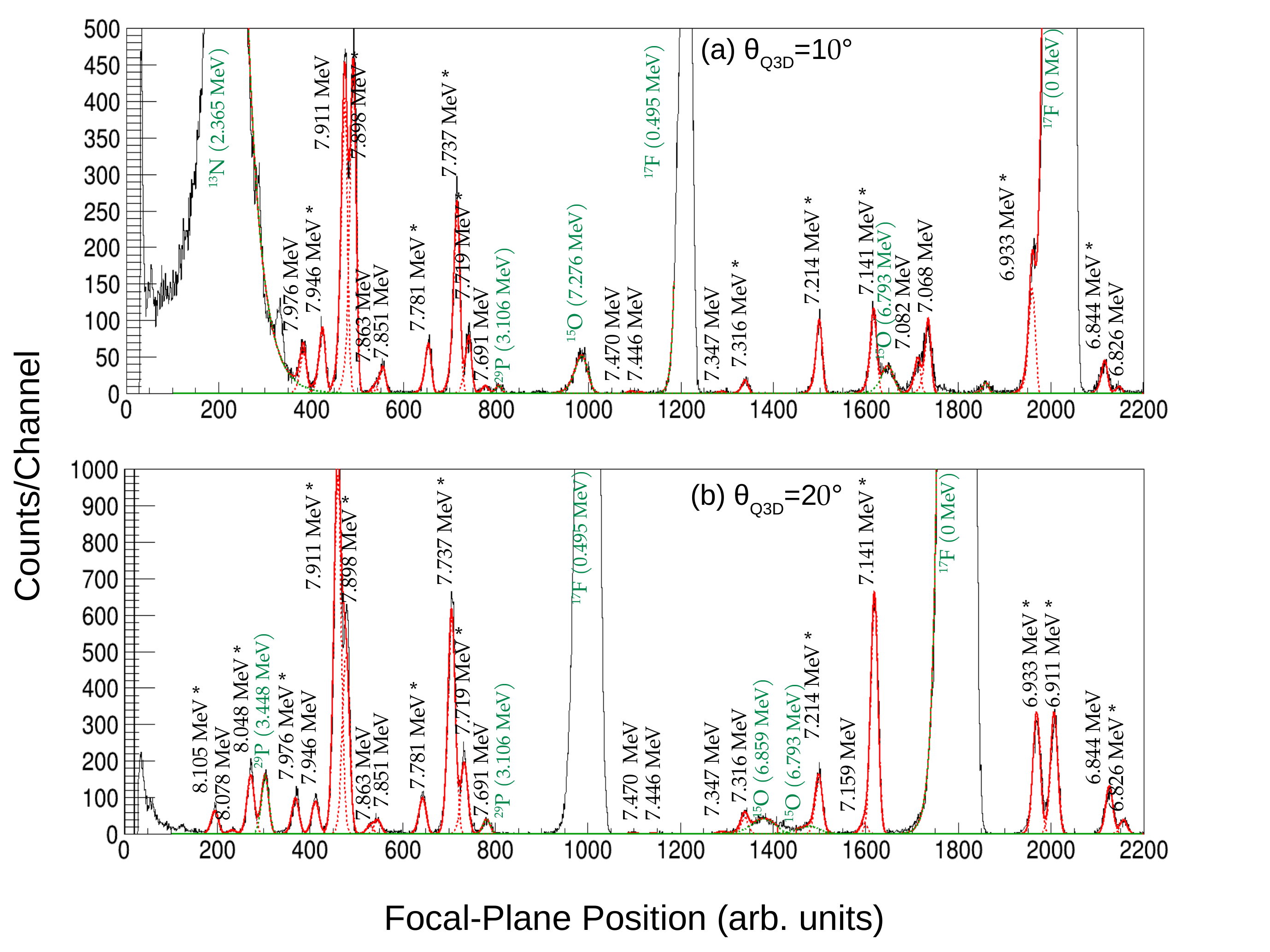}
    \caption{(Color online) Deuteron magnetic rigidity spectra at spectrometer angles of 10\degree\ (panel a)) and 20\degree\ (panel b)) corresponding to an incident charge of 1200~$\micro$C and 5768~$\micro$C, respectively. Excitation energies in $^{31}$P between $E_x = 6.8$~MeV and 8.1~MeV are covered and are reported for each state. \ph\ states labeled with an asterisk have been used in the calibration procedure. The best fit is shown as a solid line, while the individual contributions are in red/green dashed lines for $^{31}$P states and contaminant peaks, respectively.}
    \label{fig:Spectra}
\end{figure*}

After selection, deuteron spectra of the focal-plane position were obtained for each spectrometer angle. Typical spectra at 10\degree\ and 20\degree\ are shown in Fig.~\ref{fig:Spectra}. States resulting from the \ph\ reaction could be readily identified based on their kinematic shifts at different scattering angles. The origin of the remaining peaks could be associated with reactions on contaminants in the target. Since the magnetic spectrograph was tuned to compensate for the kinematic aberration associated with the \shd\ reaction, deuteron peaks induced by contaminant reactions are broader than those associated with narrow \ph\ states. Moreover the contaminant peaks exhibit a kinematic dependence which is readily distinguishable from the \ph\ excited states. Deuteron peaks associated with ($^3$He,$d$) reactions on the target components ($^{28}$Si, $^{16}$O and $^{12}$C), as well as on $^{14}$N were also identified.

The deuteron magnetic rigidity spectra were independently analysed at each detection angle using a least-squares fit of a sum of exponentially modified Gaussian functions~\cite{Heusler2010}, which proved to be a good model for the peak shape induced by the Q3D response. A common width and common exponential factor were considered as free parameters for all \ph\ states at a given angle. Contamination peaks associated with nuclei with a mass number close to $A=31$ were also fitted with the same modified Gaussian function, but with different width and exponential factors. For contaminant peaks associated to lighter nuclei the optical aberrations induced a peak shape which could not be described by a modified Gaussian function. These peaks were excluded from the fit except when peaks of interest were close by; in that case the tail of the contamination peak was modelled by a simple exponential function and considered as background (e.g. the $^{19}$F state at $E_x = 495$~keV). An energy resolution of 5~keV (FWHM) was obtained at $\theta_{lab} = 6\degree$, a significant improvement compared to the previous \shd\ study~\cite{Vernotte1990} thanks to the use of a thinner target and the better resolving power of the Q3D magnetic spectrometer.

\begin{figure}[!hbtp]
    \includegraphics[width=\linewidth]{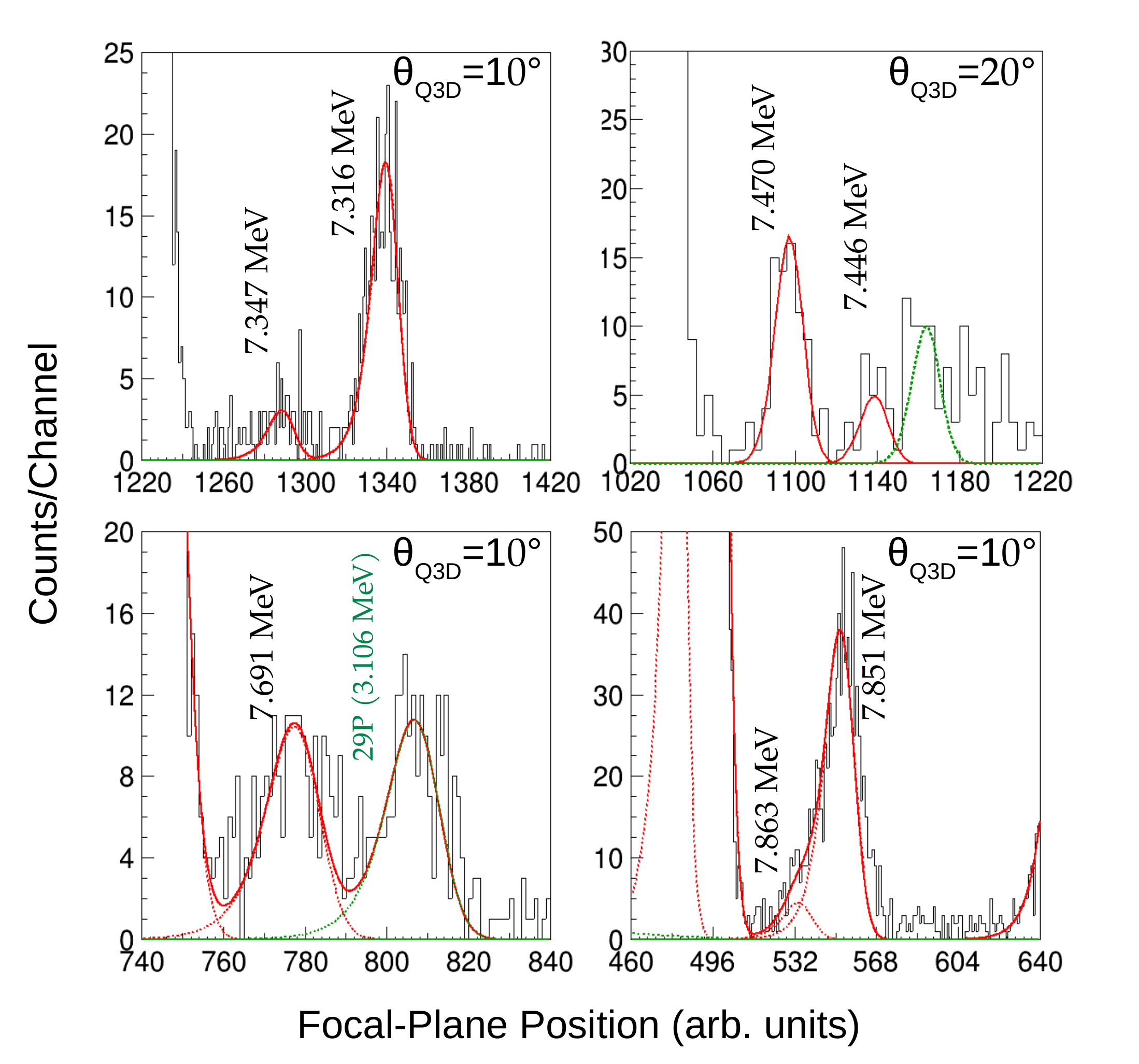}
    \caption{(Color online) Deuteron magnetic rigidity spectra showing a close-up on weakly populated unbound \ph\ levels (see text for discussion). Spectrum in the upper right panel have been re-binned with a factor 4.}
    \label{fig:zoom}
\end{figure}

Several \ph\ levels in the energy range $E_x = 6.8 - 8.1$~MeV are isolated and well populated (see Fig.~\ref{fig:Spectra}), with accurately determined excitation energies~\cite{nndc}. These levels (labeled with asterisks in Table~\ref{tab:ex} and Fig.~\ref{fig:Spectra}) were used for rigidity calibration, and a second-degree polynomial relation was obtained between the radius of curvature and the position on the focal-plane detector. The residuals between the known and determined excitation energies of \ph\ states are less than 4~keV. The calibration procedure was repeated for each spectrometer angle and the adopted excitation energies, reported in Table~\ref{tab:ex}, come from the weighted average of energies obtained at each detection angle when the state was observed. The obtained excitation energies for well-populated states in the present work are in excellent agreement with previous results. With respect to the previous \shd\ experimental study~\cite{Vernotte1990}, the excellent energy resolution achieved in the present work enabled the separation of several unresolved doublets. Moreover, several weakly populated states have been observed. Fig.~\ref{fig:zoom} shows a close-up view of the deuteron magnetic rigidity spectra for $^{31}$P levels above the proton threshold which were not observed in Ref.~\cite{Vernotte1990}. Levels above the proton threshold are discussed in detail in Section~\ref{sec:discussion}.

\begin{table*}[!htpb]
    \caption{Excitation energy, angular momentum, proton spectroscopic factor and single-particle width of \ph\ levels populated by the \shd\ reaction are reported. Italic numbers correspond to excitation energy uncertaintie. Comparison with available information from the literature is provided. \ph\ states labeled with an asterisk have been used in the calibration procedure.}
    \begin{ruledtabular}
	\begin{tabular*}{\textwidth}{@{\extracolsep{\fill} }lccclcclc}
		\multicolumn{4}{c}{Present work} & \multicolumn{3}{c}{Vernotte et al.~\cite{Vernotte1990}} & \multicolumn{2}{c}{ENSDF~\cite{nndc}} \\ 
		\cline{1-4} \cline{5-7} \cline{8-9}
		$E_x$ (keV) & $\ell_p$ & (2$J$+1)$C^2S$ & $\Gamma_{sp}$ (eV)\footnotemark[1]& $E_x$ (keV)\footnotemark[2] & $\ell_p$ & (2$J$+1)$C^2S$\footnotemark[3]& $E_x$ (keV) & $J^{\pi}$ \\
		\hline
		6826.5 \emph{ 9} & 5 & 0.061 & $1.16\times10^4$ & 6826 &   &       & 6824.2 \emph{20} & 11/2$^-$ \\
		                 &   & &        &      &   &       & 6828   \emph{ 3} &        \\
		6844.0 \emph{ 8} *& 3 & 0.034& $2.90\times10^4$ & 6843 & 3 & 0.04 & 6841.9 \emph{15} & 5/2$^-$  \\
		6911.3 \emph{10} *& 1 & 0.052&$1.58\times10^5$  & 6910 & 1 & 0.08  & 6909.6 \emph{16} & 3/2$^-$  \\
		6933.2 \emph{ 9} *& 3 & 0.085& $2.96\times10^4$ & 6932 & 2 & 0.08  & 6931.9 \emph{15} & 5/2$^+$  \\
		       & or 2  & 0.040 & $6.13\times10^4$ &  &  &  &  &   \\
		7068.5 \emph{ 9} & 3 & 0.041& $3.06\times10^4$ & 7068 & 3 & 0.04 & 7073   \emph{ 4} & 5/2$^-$, 7/2$^-$\\
		                 &   &   &     &      &   &       & 7073   \emph{ 6} & 1/2$^+$, 3/2$^+$ \\
	    7082.4 \emph{14} & 1  & 0.004 & $1.65\times10^5$  & 7081 &   &       & 7079.9 \emph{19} & 3/2$^-$, 5/2$^+$ \\
	                     &    &      &  &      &   &       & 7084.0 \emph{17} & 3/2$^+$, 5/2, 7/2$^+$\\
	                     &   &     &   &      &   &       & 7117.7 \emph{10} & 9/2$^+$ \\
	    7140.7 \emph{ 8} *& 0 & 0.099 & $1.61\times10^5$ & 7139 & 0 & 0.22  & 7141.1 \emph{18} & 1/2$^+$ \\
	    7159.7 \emph{16} & 1 & 0.002& $1.60\times10^5$ &      &   &       & 7158   \emph{ 5} &       \\
		       & or 3  & 0.003 & $3.12\times10^4$ &  &  &  &  &   \\
	    7214.4 \emph{ 8} *& 1 & 0.020& $1.61\times10^5$  & 7214 & 1 & 0.032 & 7214.3 \emph{20} & 1/2$^-$, 3/2$^-$ \\ 
	                     &   &     &   &      &   &       & 7313.7 \emph{16} & 1/2$^+$, 3/2$^+$ \\
	    7316.1 \emph{ 9} *& 3 & 0.0075& $2.44\times10^{-36}$ & 7314 & 3 & 0.016 &   7314    \emph{ 4}            &   $5/2^-, 7/2^-$          \\
	    7347.1 \emph{12} & 1 & 0.0007& $2.78\times10^{-17}$ &      &   &       & 7346   \emph{ 6} & 3/2$^-$, 5/2$^-$ \\
	            & or 2  & 0.0012 & $6.60\times10^{-19}$ &  &  &  &  &   \\
	                     &   &     &   &      &   &       & 7356 \emph{16} &  \\
	    7445.7  \emph{29}& 2  & 0.0007& $6.71\times10^{-8}$ &      &   &       & 7441.4 \emph{10} & (3/2 to 9/2) \\
	            & or 3  & 0.0006 & $1.19\times10^{-9}$ &  &  &  &  &   \\
	                     & &  &        &      &   &       & 7442.3 \emph{ 3} & 11/2$^+$ \\
	    7470.4 \emph{22} & 3  & 0.0008& $1.60\times10^{-8}$ &      &   &       & 7466   \emph{ 2} & 5/2$^-$, 7/2$^-$, 9/2$^-$\\
	            &         &   &        &      &   &       & 7572             &       \\
	    7690.9 \emph{10} & 3 & 0.006& $1.53\times10^{-3}$ &      &   &       & 7687.2 \emph{20} &      \\
	    7719.4 \emph{ 8} *& 3 & 0.045& $3.34\times10^{-3}$ & 7718 &   &       & 7718 \emph{4}                &     \\
	    7737.3 \emph{ 8} *& 3 & 0.114& $6.89\times10^{-3}$ & 7736 & 3 & 0.16  &     7736  \emph{4}   & $5/2^-, 7/2^-$      \\
	    7781.1 \emph{ 8} *& 1 & 0.015& $16.25$ & 7780 & 1 & 0.02 & 7779   \emph{ 1} & 3/2$^-$ \\
	                     &   &     &   &      &   &       & 7825   \emph{ 9} &      \\
	    7851.2 \emph{ 8} & 1 & 0.009& $55.75$ & 7855 &   &       & 7852   \emph{ 4} & 1/2,3/2,5/2$^+$ \\
	            & or 2  & 0.0114 & $2.56$ &  &  &  &  &   \\
	    7863.8 \emph{14} & 3 & 0.004& $0.074$  &      &   &       & 7859.8 \emph{ 4} & 11/2$^-$  \\
	    7898.0 \emph{ 8} *& 1 & 0.115& $112.6$  & 7900 & 1 & 0.32  & 7896   \emph{ 1} & 1/2$^-$ \\
	    7911.5 \emph{ 8} *& 3 & 0.190& $0.215$  & 7913 & 3 & 0.12  &     7913   \emph{4}             &     $5/2^-, 7/2^-$   \\
	    7946.2 \emph{ 8} *& 2 & 0.033& $8.711$  & 7949 & 2 & 0.04  & 7945   \emph{ 1} & 3/2$^+$, 5/2$^+$ \\
	    7976.4 \emph{ 8} & 2 & 0.023& $16.122$  & 7980 & 2 & 0.032 &      7980            &            \\
	                     &   &   &     &      &   &       & 7994   \emph{ 6} & 5/2$^-$      \\    
	                     &   &    &    &      &   &       & 8031   \emph{ 1} & 5/2$^+$      \\  
	    8048.4 \emph{11} *& 1 & 0.034& $733.5$  & 8051 & 1 & 0.04  & 8048   \emph{ 1} & 3/2$^-$   \\
	    8078.0 \emph{17} *& 1 & 0.004& $975.3$  & 8080 &   &       & 8077.0 \emph{ 4} & 11/2$^-$  \\
	    8104.9 \emph{15} *& 2 & 0.023& $68.11$  & 8107 & 2 & 0.018 & 8104   \emph{ 1} & 5/2$^+$   \\
	\end{tabular*}
	\end{ruledtabular}
	\footnotetext[1]{Reduced single-particle proton width $\gamma^2_{s.p.}$ is given for bound states up to $S_p=7296.55(2)$~keV.}
	\footnotetext[2]{Excitation energies are given with an uncertainty of $\pm4$~keV.}
	\footnotetext[3]{The protons are assumed to be transferred to $2p_{3/2}$, $1d_{3/2}$ and $1f_{7/2}$ orbitals for $\ell= 1, 2 \text{ and }3$ transitions, respectively.}
	\label{tab:ex}
\end{table*}

\section{Results}
\label{sec:results}
\subsection{Angular distributions and DWBA analysis}
\label{sec:res_A}
\begin{figure*}[!htpb]
    \includegraphics[width=\textwidth]{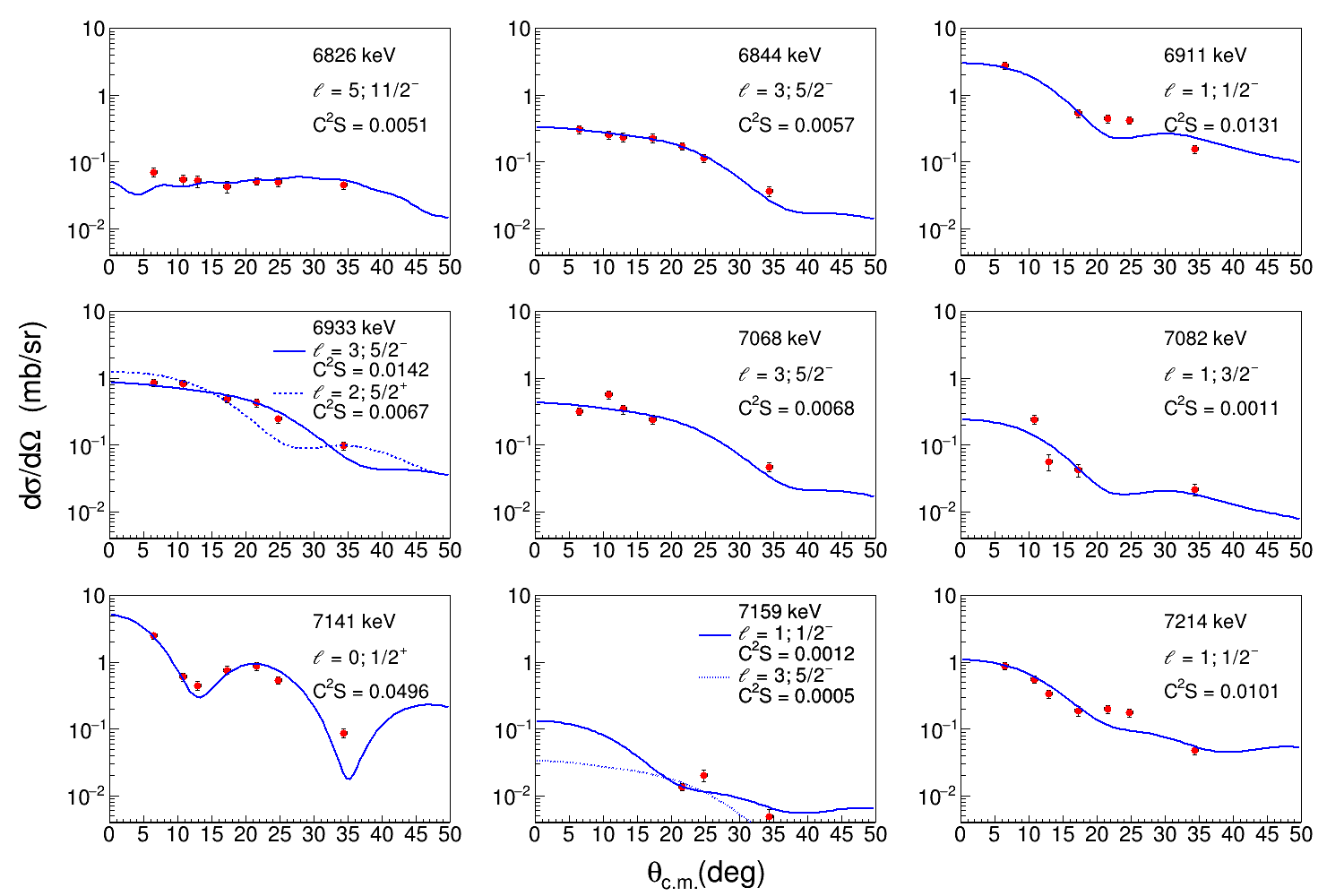}
    \caption{(Color online) Angular distributions of \ph\ bound states populated in the \shd~reaction. Curves represent finite-range DWBA calculations normalized to the data. The error bars are typically smaller than the data points.}
    \label{fig:angular distributions bound}
\end{figure*}

\begin{figure*}
    \centering
    \includegraphics[width=\textwidth]{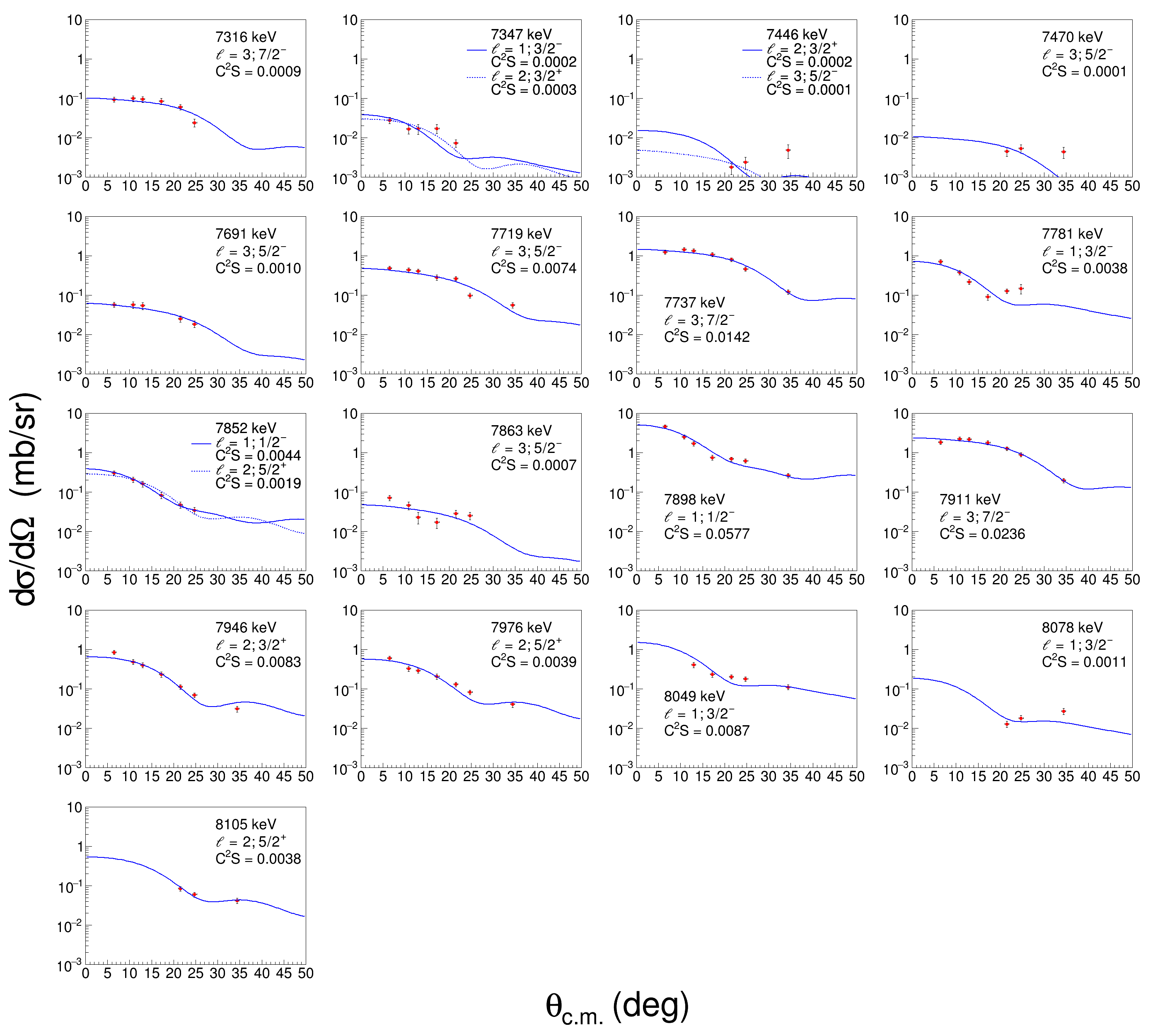}
    \caption{(Color online) Angular distributions of \ph\ proton-unbound states populated in the \shd~reaction. Curves represent finite-range DWBA calculations normalized to the data. The error bars are typically smaller than the data points. 
    }
    \label{fig:angular distributions unbound1}
\end{figure*}

The differential cross sections corresponding to states populated in the \ph\ reaction were calculated from the deuteron yield determined at each spectrometer angle after a proper normalisation taking into account the areal density of \sit\ atoms in the target, the solid angle and the accumulated charge corrected for the electronic dead time. The differential cross-sections for twenty-seven \ph\ levels are shown in Fig.~\ref{fig:angular distributions bound} and in Fig.~\ref{fig:angular distributions unbound1} for states below and above the \sit+p threshold ($S_p=7296.55(2)$~keV~\cite{nndc}), respectively. Finite-Range Distorted Wave Born Approximation (FR-DWBA) calculations have been performed with the {\sc Fresco} code~\cite{FRESCO} and the best fit to the data is also plotted.

The optical potential parameters for the entrance channel come from an experimental study of the same transfer reaction at the same bombarding energy~\cite{Vernotte1982}, and the parameters for the exit channel come from set F of the Daehnick {\it et~al.} global deuteron potentials~\cite{Daehnick1980}. The finite-range calculations have been performed using the $\langle^3$He$\mid$d$\rangle$ overlap obtained with the Green's function Monte Carlo method using the Argonne $\nu_{18}$ two-nucleon and Illinois-7 three-nucleon interactions~\cite{Brida2011}. The wave-function describing the $^{30}$Si$+p$ relative motion was calculated by adjusting the depth of a Woods-Saxon potential to reproduce the known separation energy of each \ph\ state. The optical potential parameters used in the present analysis are reported in Table~\ref{tab:pot}. A very good agreement is obtained between normalized FR-DWBA calculations and the experimental differential cross-sections (see Fig.~\ref{fig:angular distributions bound}) which supports a single-step direct mechanism for the proton transfer reaction populating \ph\ excited states.
\begin{table*}[t]
    \caption{Potential parameters used in DWBA calculations}
    \begin{ruledtabular}
   \begin{tabular*}{\textwidth}{ccc c cc c c cc c c}
    \noalign{\smallskip}
   Channel & $V_{WS}$ \newline (MeV) & $r_r$ \newline (fm) & $a_r$ \newline (fm) & $W_V$ \newline (MeV) & $W_D$ \newline (MeV) & $r_i$ \newline (fm) & $a_i$ \newline (fm) & $V_{s.o.}$ \newline (MeV) & $r_{s.o.}$ \newline (fm) & $a_{s.o.}$ \newline (fm) & $r_C$ \newline (fm)\\
   \hline
    $^{30}$Si + $^{3}$He & 189.8 & 1.150 & 0.669  & 24.0 &      & 1.495 &  0.886 & & & & 1.4 \\
    $^{31}$P + $^{2}$H   &  85.7 & 1.170 & 0.755  & 0.90 & 12.0 & 1.325 &  0.749 & 3.28 & 1.07 & 0.66 & 1.3 \\
    $^{2}$H + p \footnotemark[1] &  179.94 & 0.540 & 0.680 &   &   &   &   &   &   &   & 1.25\\
    $^{30}$Si + p        &  adjusted & 1.250 &  0.650 &  &  &  &  & 6.25 & 1.25 & 0.650 & 1.25 \\
    \end{tabular*}
    \end{ruledtabular}
    \footnotetext[1]{The parametrization of overlaps calculated in Ref.~\cite{Brida2011} adds a Gaussian component to the Wood-Saxon potential: $V_{WS} \times \left\{\frac{1}{1+ \exp[(r-r_r)/a_r]} - \beta \exp[-(r/\rho)^2] \right\} $  where $\beta = 1.13$ and $\rho = 0.64$}
    \label{tab:pot}
\end{table*}

The normalization factor obtained from fitting the theoretical angular distributions to the data is related to the proton spectroscopic factor $C^2S_p$. For bound \ph\ levels (positive binding energy) the spectroscopic factors are directly deduced from the normalization procedure since the determination of the form factor needed by the DWBA calculation is fairly simple in that case. For unbound \ph\ states the FR-DWBA calculations were performed for several positive binding energies down to 10~keV, and the spectroscopic factor was then linearly extrapolated to the binding energy (negative) of the state under consideration~\cite{Ser17}. We estimate that the finite-range calculations introduce no more than 5\% uncertainty based on a similar extrapolation performed with Zero-Range DWBA calculations.

The values of the proton spectroscopic factor are displayed in Table \ref{tab:ex}. These values, obtained using finite range calculations, are lower by $25\%-40\%$ than those obtained in the zero-range analysis of \ver. This is in agreement with theoretical calculations where a reduction factor of 0.77 is found, due to finite-range considerations~\cite{Bassel1966}. 

The uncertainty on the spectroscopic factor comes from ($i$) the optical potentials describing the elastic channels, ($ii$) the binding potential used for the calculation of the relative $p+^{30}$Si wave function, and ($iii$) the experimental uncertainties associated to the cross section determination. The uncertainties associated to the optical potentials for (\nuc{3}{He},d) transfer reactions are commonly reported in the literature to be about 30\% of the nominal value~\cite{Vernotte1990}. The uncertainty arising from the poorly constrained geometry of the binding potential has been investigated with a Monte-Carlo approach where the radius and diffuseness of the associated Wood-Saxon well were independently sampled according to a Gaussian distribution with a full width at half maximum of 25\% (radius) and 35\% (diffuseness) with respect to the nominal values (see Table~\ref{tab:pot}). This study was performed for $s$-, $p$-, $d$- and $f$-waves resulting in a 30\% uncertainty on the spectroscopic factor. The experimental and statistical fit errors are less than 7\% for the majority of levels observed, while they represent 11\% for low populated states at $E_x=7082$-, 7159-, 7347-, and 8078-keV. For states of astrophysical interest at  $E_x=7446$- and 7470-keV, the errors are 28\% and 17\% respectively.

\subsection{Proton width and resonance strength}
For unbound states the proton width $\Gamma_p$ is expressed as the product of the proton spectroscopic factor and the single-particle proton width: $\Gamma_p = C^2S_p\times\Gamma_{s.p}$. The single-particle width is calculated with the formula~\cite{iliadisbook}:
\begin{equation}
  \Gamma_{s.p.} = \frac{\hbar^2 s}{2 \mu} P_\ell(E_r, s) |R(s)|^2,
  \label{eq:spwidth}
\end{equation}
where $\mu$ is the reduced mass of the $p+^{30}$Si system, $P_l(E_r,s)$ is the penetrability of the Coulomb and centrifugal barriers for transferred angular momentum $\ell$, and $R(s)$ is the radial part of the $p+^{30}$Si wave function. Both the penetrability and the radial part of the wave function should be evaluated at a radius $s$ where the latter approaches an asymptotic behaviour. In the present work the calculations were performed with the radius $s=7$~fm because it allowed the radial part of the wave function to fulfill the previous condition. The code {\sc Dwuck4} was used to calculate the single-particle width since it provides an easy way to solve the Schr\"odinger equation for the $p$+\sit\ scattering state. The same interacting potential as the one used for determining spectroscopic factors with {\sc Fresco} was used, and the depth of its central volume part was varied until a resonance was obtained at the right energy~\cite{Iliadis1997}.

The uncertainty on the proton width results from the uncertainties on the spectroscopic factor and the single-particle wave function. Since both $C^2S_p$ and $\Gamma_{s.p.}$ strongly depend on the geometry of the binding potential a certain degree of anti-correlation is expected between their uncertainties. The Monte-Carlo technique presented in section~\ref{sec:res_A} was used to calculate the single-particle width for each radius and diffuseness combination defining the geometry of the potential well. A strong correlation between the spectroscopic factor and the single-particle width was observed resulting in an uncertainty on the resulting proton width $\Gamma_p$ smaller than 1\%. The impact of these correlations on the extraction of the proton widths will be explored in a forthcoming publication. The final uncertainty on the proton width is then given by the uncertainty on the spectroscopic factor due to the optical potentials describing the entrance and exit channels, e.g. $\approx30\%$. 

Once the proton width is calculated the resonance strength $\omega\gamma$ can be evaluated. The resonance strength is defined as~\cite{iliadisbook}: 
\begin{equation}
    \omega\gamma = \frac{(2J+1)}{(2j+1)(2J_{\nuc{30}{Si}}+1)}\frac{\Gamma_p\Gamma_\gamma}{\Gamma_{tot}},
    \label{eq:omega_gamma}
\end{equation}
where $J, j$ and $J_{\nuc{30}{Si}}$ are the spins of the resonance, the proton and the target nucleus, respectively.  $\Gamma_{tot}=\Gamma_p+\Gamma_\gamma$ is the total width, the sum of the proton partial width ($\Gamma_p$) and the $\gamma$-ray partial width ($\Gamma_\gamma$) since these two channels are the only ones open in the energy range of our study. For excitation energies near the proton threshold, the $\gamma$-decay channel dominates ($\Gamma_p \ll \Gamma_\gamma$). Given that $j=1/2$ and $J_{\nuc{30}{Si}}=0$, the resonance strength is thus well approximated as $\omega\gamma=0.5(2J+1)\Gamma_p$. Values of the proton widths and resonance strengths are reported in Table~\ref{table:resonances}.

\subsection{Discussion}
\label{sec:discussion}
We present here the spectroscopic properties of the \ph\ states above the proton threshold ($S_p=7296.55(2)$~keV~\cite{nndc}). Excitation energies are used to derive resonance energies using the relation $E_r^{c.m.} = E_x - S_p$, and the uncertainty associated to the resonance energy is dominated by the one on excitation energies. Reported excitation energies are from this work, or from the literature when the \ph\ states have not been observed in the present data.

\subsubsection{$E_x=7313.7\pm1.6$ keV}
This level corresponds to a resonance energy \Ercm{17.2}. The $J^\pi = 1/2^+, 3/2^+$ assignment reported in ENSDF~\cite{nndc} for this state comes from the study of the $^{29}$Si($^{3}$He,p)\ph\ transfer reaction~\cite{al_jadir1980}. There is no indication of this level in our experimental data.

\subsubsection{$E_x=7316.2\pm0.9$ keV}
This level corresponds to a resonance energy \Ercm{19.6}. The analysis of the angular distribution from the present data is best described by a $\ell=3$ transferred orbital angular momentum, thus corresponding to a 5/2$^-$ or 7/2$^-$ assignment. This is in line with the results obtained from a previous study of the \shd\ transfer reaction~\cite{Vernotte1990}. Note that the latter work observed this state at $E_{x}=7314\pm4$~keV, which is in agreement with our value within error bars. 

\subsubsection{$E_x=7347.0\pm1.2$ keV}
This level corresponds to a resonance energy \Ercm{50.5}. This state has a reported energy $E_x=7346\pm6$~keV and spin-parity $J^\pi=3/2^-,5/2^-$ in ENSDF~\cite{nndc} based on the study of the $^{29}$Si($^{3}$He,p)\ph\ reaction~\cite{al_jadir1980}. This state is weakly populated in the present data, and its associated angular distribution is best described by $\ell=1$ or $\ell=2$ momentum transfer, giving similar reduced chi-square when only the three most forward angles are considered. Given the established spins and negative parity assignment from Ref.~\cite{al_jadir1980}, and considering the parity conservation in the present transfer reaction, this leads to $\ell=1$ and $J^\pi=3/2^-$. 

\subsubsection{$E_x=7356$ keV}
This level corresponds to a resonance energy of \Ercm{59.5}. No spin-parity assignment has been reported in the literature. This state has been populated by the \reaction{33}{S}{d}{$\alpha$}{31}{P} reaction~\cite{Teterin1974}, however the $\alpha$-particle peak associated to this state appears as a weak shoulder in a contamination peak. This prevented an accurate determination of the excitation energy, and indeed no associated uncertainty is given in that work. Although not cited in the ENSDF compilation~\cite{nndc}, the study of the \reaction{29}{Si}{\nuc{3}{He}}{p}{31}{P} reaction~\cite{Moss1969} reports a state at  $E_x=7356\pm9$~keV. However this should be considered as tentative since the author gives the energy between brackets (see his Table 1). Interestingly, in a subsequent \reaction{29}{Si}{\nuc{3}{He}}{p}{31}{P} reaction study~\cite{al_jadir1980} there is no indication of the $E_x=7356$~keV state, and it was instead associated to the $E_x=7347\pm6$~keV state. This level has not been populated in the present experiment in line with its non-observation in all of the other ($^3$He,d) studies. We conclude that the existence of the $E_x= 7356$~keV state is not firmly established and we thus do not further consider it in this work.
 
\subsubsection{$E_x=7441.4\pm1.0$ keV and $E_x=7442.3\pm0.3$ keV}
\label{sec:149}
This doublet that is reported in Ref.~\cite{nndc} corresponds to resonance energies of \Ercm{144.9} and \Ercm{145.8}. The low energy component of the doublet was first observed using the \reaction{27}{Al}{$\alpha$}{$\gamma$}{31}{P} reaction~\cite{DEVOIGT1971} and a spin assignment $J=(3/2-9/2)$ was derived~\cite{nndc}. This state mainly decays through two $7/2^+$ levels thus adding a further constraint for its parity which is most likely positive in case of a $J=3/2$ assignment. The other component of the doublet has been observed in several experimental studies~\cite{Twin1974,jenkins2005,ionesco2006} pointing to a $J^{\pi}=11/2^+$ assignment.

In the present experiment, we observe a weakly populated state at the three angles $\theta_{Q3D}\geq20\degree$, with a weighted average excitation energy of \Ex{7445.7\pm2.8}. Since the present transfer reaction is better matched for low transferred angular momenta, we assume that the level observed corresponds to the low spin component of the doublet. The spin-parity assignment discussed previously corresponds to either $\ell = 2, 3$ or $4$. However the limited angular range of the angular distribution and the lack of data at forward angles does not allow to discriminate between the different possible transferred relative angular momenta. We therefore consider $\ell=2$ or $\ell=3$ transfers; both DWBA calculations are reported in Fig.~\ref{fig:angular distributions unbound1}, and the impact of these possibilities will be discussed in the reaction rate section (Sec.~\ref{sec:rate}).

\subsubsection{$E_x=7470.5\pm2.3$ keV}
This level corresponds to a resonance energy of \Ercm{174.0}. A deuteron peak is unambiguously associated to this state at $\theta_{Q3D}=20\degree, 23\degree$ and $32\degree$. We associate this level to the one reported at $E_x=7466\pm2$~keV in the literature. The $E_x=7466$~keV state was first observed using the \reaction{27}{Al}{$\alpha$}{$\gamma$}{31}{P} reaction~\cite{DEVOIGT1971}, and later with the \reaction{28}{Si}{$\alpha$}{p$\gamma$}{31}{P}~\cite{Twin1974} and \reaction{29}{Si}{\nuc{3}{He}}{p}{31}{P}~\cite{al_jadir1980} reactions. Based on these experimental studies the spin-parity of this state is restricted to $J^{\pi} = (7/2, 9/2)^-$~\cite{ENDT1990}. Such spin-parity assignment corresponds to an $\ell=3$ or $\ell=5$ angular momentum transfer for the \shd\ reaction. The spectroscopic factor was extracted for the lower angular momentum transfer in order to maximize the strength of this resonance.

\subsubsection{$E_x=7572$ keV}
This level corresponds to a resonance energy of \Ercm{276}. It has only been observed in the experimental study of the \reaction{33}{S}{d}{$\alpha$}{31}{P} reaction~\cite{Teterin1974}, no spin-parity is given. There is no evidence of this state in the present study, in agreement with the results from other \shd\ study~\cite{Vernotte1990}.

\subsubsection{$E_x=7691.1\pm1.0$ keV}
This level corresponds to a resonance energy \Ercm{394.6}. This state has been unambiguously observed in the present experiment at $\theta_{Q3D}=6\degree, 10\degree$ and $12\degree$. At higher angles, the state at $E_x=7691$~keV overlaps with the \nuc{29}{P} state at $E_x=3105$~keV populated by the \reaction{28}{Si}{\he}{d}{29}{P} reaction. For large Q3D detection angles, the angular distribution of the \ph\ state at $E_x=7691$~keV was thus obtained after subtracting the contribution from the \nuc{29}{P} level. Such contribution was estimated using the angular distribution obtained in the \reaction{28}{Si}{\he}{d}{29}{P} study~\cite{dykoski1976} performed at the same incident energy as the present experiment. The angular distribution of the $E_x=7691$~keV state of \ph\ is best described by a $\ell=3$ orbital angular momentum transfer, giving $J^\pi = 5/2^-$ or $7/2^-$. This level was not observed in the work of Ref.~\cite{Vernotte1990} because of the presence of the \nuc{29}{P} contamination at all angles. 

We associate the present level with the \Ex{7687.2\pm2.0} state which has been observed in a single experiment using the \spg\ reaction~\cite{deneijs1975}. Based on the $\gamma$-ray feeding and decay properties, $J \leq 9/2$ is suggested~\cite{nndc}, which is in agreement with the present spin-parity assignment.

\subsubsection{$E_x=7719.5\pm0.8$ keV}
This level corresponds to a resonance energy of \Ercm{423.0}. It has been observed using the \reaction{29}{Si}{\nuc{3}{He}}{p}{31}{P}~\cite{al_jadir1980},  \reaction{30}{Si}{d}{n}{31}{P}~\cite{UZUREAU1976} and \shd~\cite{Vernotte1990} transfer reactions. No spin-parity assignment could be established because of the proximity of the \Ex{7737} state forming an unresolved doublet in the aforementioned experiments. The resolution obtained in the present experiment allowed to extract the angular distribution of this state at all Q3D angles. The analysis of this distribution shows a $\ell=3$ orbital angular momentum transfer pattern leading to $J^\pi = 5/2^-$ or $7/2^-$. A recent \spg\ direct measurement~\cite{Dermigny2020} suggests $J^{\pi} =(3/2, 5/2)^-$ based on the combination of constrains coming from the $\gamma$-ray transitions and the \reaction{29}{Si}{\nuc{3}{He}}{p}{31}{P}\ reaction~\cite{al_jadir1980}. From the combination of all existing spin-parity assignments, we suggest that the \Ex{7719.5} state has $J^\pi=5/2^-$.

The resonance in the \spg\ reaction is observed at \Ercm{421.9\pm0.3} with a strength  $\omega\gamma=(1.14\pm0.25)\times10^{-4}$~eV~\cite{Dermigny2020} . Using the value of the proton width deduced from the spectroscopic factor $(2J+1)C^2S=0.045 $, and under the reasonable approximation $\Gamma_{p}\ll \Gamma_{\gamma}$ at this energy, we find the strength of this resonance  $\omega\gamma=(7.43\pm2.23)\times10^{-5}$~eV. For the reaction rate calculation we favor the direct measurement of the strength of this resonance.

\subsubsection{$E_x=7737.3\pm0.8$ keV}
This level corresponds to a resonance energy of \Ercm{440.8}. It has only been observed through the \shd\ reaction \cite{Vernotte1990} at \Ex{7736\pm4}, forming an unresolved doublet with the \Ex{7719} state. This study suggested a $\ell=3$ transferred angular momentum for the doublet angular distribution. The present analysis of the angular distribution of this level obtained at all angles supports the $\ell=3$ value.

\subsubsection{$E_x=7781.1\pm0.8$ keV}
This level corresponds to a resonance energy of \Ercm{484.6}. It has been observed by several transfer reactions~\cite{Vernotte1990, al_jadir1980, UZUREAU1976} and in the direct measurement of the \spg\ reaction~\cite{deneijs1975, Dermigny2020}. These direct measurement experiments established a $J^\pi=3/2^-$ spin-parity assignment~\cite{nndc}. The angular distribution obtained in the present work at all angles corresponds to a transferred angular orbital momentum of $\ell=1$ in line with the known spin-parity assignment.

The resonance in the \spg\ reaction is observed at \Ercm{486.2\pm0.2} with a strength  $\omega\gamma=(0.188\pm0.014)$~eV~\cite{Dermigny2020}. Using the value of the proton width deduced from the spectroscopic factor $(2J+1)C^2S=0.016 $, and under the reasonable approximation at this energy that $\omega\gamma=0.5(2J+1)\Gamma_p$, we find a strength of  $\omega\gamma=0.12\pm0.04$~eV for this resonance. For the reaction rate calculation we favor the direct measurement of the strength of this resonance.

\subsubsection{\Ex{7825\pm9}}
This level corresponds to a resonance energy of \Ercm{528.5}. It has only been observed in the study of the \reaction{29}{Si}{\nuc{3}{He}}{p}{31}{P} reaction~\cite{Moss1969} even though it has not been confirmed by the study of the same transfer reaction at a different beam energy~\cite{al_jadir1980}. In the present work there is an indication of a deuteron peak at the expected energy at a detection angle of 10\degree. However, this indication at a single angle is not enough to conclude for a positive observation of this state. We therefore discard this level in the rest of the present study.

\subsubsection{\Ex{7851.4\pm0.8}}
This level corresponds to a resonance energy of \Ercm{554.9}. It has been observed in the $^{29}$Si($^{3}$He,p)\ph\ experiment~\cite{al_jadir1980} at \Ex{7856 \pm 6 } with $J= 1/2 - 5/2$, and with the \reaction{32}{S}{d}{\he}{31}{P}\ reaction~\cite{VERNOTTE1999} at \Ex{7851\pm5}. This state was also observed at \Ex{7855 \pm 4} in a previous \shd\ experiment~\cite{Vernotte1990}, but the corresponding angular distribution could not be reproduced by DWBA calculations. This is probably due to the presence of the tail of the strongly populated state at~\Ex{7898}, together with the existence of an unresolved state at \Ex{7863}. 

In the present study, the spectrum in this excitation energy region is better described by two close components \Ex{7851.4\pm0.8} and \Ex{7863.4\pm1.4} ($\chi^2$/ndf = 0.95 against $\chi^2$/ndf =1.52 for one-component fit). The angular distribution of the \Ex{7851.4\pm0.8} state could be extracted and is well described by a DWBA calculation corresponding to either $\ell = 1$ or $\ell = 2$.

The gamma width of this resonance is deduced from a ($\gamma$,$\gamma$') experimental study~\cite{SHIKAZONO1972} and estimated to $\Gamma_{\gamma}~=~0.7\pm0.1$~eV (for $\ell=1$). This value is of the same order of magnitude than the proton width, thus the approximation $\omega\gamma=0.5(2J+1)\Gamma_p$ starts to be no longer justified for these excitation energies and the exact strength formula (Eq. \ref{eq:omega_gamma}) is used in that case.

\subsubsection{\Ex{7863.4\pm1.4}}
This level corresponds to a resonance energy of \Ercm{566.9}. Despite its weak population, it has been observed at 5 angles. The only state that could be associated to our observation would be the state at \Ex{7859.8 \pm 0.4} reported in the \reaction{24}{Mg}{\nuc{16}{O}}{2$\alpha$p$\gamma$}{31}{P}\ reaction~\cite{ionesco2006} with a spin parity assignment of $J^{\pi} = 11/2^-$. 

The angular distribution for this state can be described by either a $\ell=0$ or $\ell\ge3$ transferred orbital momentum with similar statistical likelihoods. However the very weak population of this state may suggest a bad matching condition for the transfer reaction, which would favor large transferred angular momenta.

In addition, this resonance has not been reported in the low-energy study of the \spg\ reaction~\cite{KUPERUS1959} although two $\ell=1$ resonances were observed in its close vicinity at \Ercm{484.6} and \Ercm{601.5}. This could suggest that the contribution of the present resonance is suppressed by the Coulomb and centrifugal barriers, which would indicate a rather high spin.

We therefore consider for this state $\ell=3$, and this must be considered as maximizing its contribution to the reaction rate.

\subsubsection{\Ex{7898.0\pm0.7}}
This level corresponds to a resonance energy of \Ercm{601.3}. It has been observed in several transfer reactions~\cite{al_jadir1980,UZUREAU1976, Vernotte1990}, and through $\gamma$-ray studies of \ph\ decays~\cite{KUPERUS1959, deneijs1975, SHIKAZONO1972}. The direct \spg\ measurements established a $J^\pi=1/2^-$ spin-parity assignment and total width of $\Gamma=68\pm9$~eV~\cite{KUPERUS1959}. This state has been well populated at all angles in the present work and the analysis of its angular distribution is well described by a $\ell=1$ transferred angular momentum, in agreement with the accepted spin-parity assignment. 

The proton width calculated in the present work is $\Gamma_p = 68$~eV and represents the main contribution to the total width $\Gamma=68\pm9$~eV~\cite{KUPERUS1959}. The resonance strength is thereby proportional to $\Gamma_{\gamma}$ which can not be estimated from the present experimental data. We therefore use the known resonance strength $\omega\gamma=1.95\pm10$~eV~\cite{PAINE1979} for the reaction rate calculations.

\subsubsection{$E_x>7900$~keV}
For levels lying at excitation energies above \Ex{7900}, the resonance strength can not be estimated using the sole proton width which is determined in the present work, because of the competition with the $\gamma$-decay. We therefore make a common discussion for these states emphasizing the differences with the results from the experimental \shd\ study performed by Vernotte~{\it et al.}~\cite{Vernotte1990}. Spectroscopic data for these levels are summarized in Table~\ref{tab:ex} along with the single-particle proton widths.

There are 8 \ph\ states reported in ENSDF~\cite{nndc} up to \Ex{8200}, 6 of them have been populated and studied in the work of Vernotte~{\it et al.}. The present experiment also populates them and the results of the subsequent DWBA analysis are in excellent agreement in terms of both the transferred angular momentum and the proton spectroscopic factors. The only exception concerns the relatively weakly populated \Ex{7976} state, for which the angular distribution could be described by either a $\ell=2$ or $3$ angular momentum transfer. In the present study the angular distribution is better described by $\ell=2$. This better assignment is probably related to the fact that the \Ex{7976} state is well isolated in the present data, while it is on top of a relatively large background of unknown origin in Ref.~\cite{Vernotte1990}. 

Levels reported at $E_{x}=7994$ and $8031$~keV have not been observed in the present experiment nor in Ref.~\cite{Vernotte1990}. Concerning the \Ex{7994} level, it has only been observed using the $^{29}$Si($^{3}$He,p)\ph\ reaction~\cite{al_jadir1980} and the associated proton peak is strongly contaminated. The existence of the \Ex{7994} state may be questionable.

\section{The \spg\ reaction rate}
\label{sec:rate}

\subsection{Method}
The thermonuclear reaction rate per pair of particles is defined as \cite{iliadisbook}:
\begin{equation}
    \langle\sigma\nu\rangle = \left(\frac{8}{\pi\mu}\right)^{1/2} \frac{1}{(kT)^{3/2}} \int_{0}^{\infty} E\sigma(E)e^{-E/kT}dE,
    \label{eq:rate_general}
\end{equation}
where $\mu$ is the reduced mass of the $^{30}$Si$+p$ system, $k$ the Maxwell-Boltzmann constant, $T$ the temperature, and $\sigma(E)$ the nuclear reaction cross section at energy $E$. For the \spg~reaction, the resonances of interest are narrow and isolated and the thermonuclear reaction rate becomes~\cite{iliadisbook}:
\begin{equation}
    \langle\sigma\nu\rangle = \left(\frac{2\pi}{\mu kT}\right)^{3/2}\hbar^2 \sum_i  (\omega\gamma)_i e^{-E/kT}dE,
    \label{eq:rate_narrow}
\end{equation}
where $(\omega\gamma)_i$ is the strength of each resonance as defined in Eq.~\ref{eq:omega_gamma}. 

The reaction rates have been estimated numerically using the {\sc RatesMC} code relying on a Monte Carlo method presented in Ref.~\cite{ILIADIS2010251}. In this procedure, the resonance energies $E_r^{c.m.}$ are sampled from a Gaussian distribution, with an uncertainty estimated from the excitation energies obtained in the present work. The resonance strengths $\omega\gamma$, or the proton and $\gamma$-ray partial widths, are sampled from a log-normal distribution  characterized by the associated experimental uncertainties. For the states observed in the present work, except for those cases when either the total or $\gamma$-ray width are determined, the $1\sigma$ strength uncertainty originates from the proton width uncertainty estimated from the quadratic sum of the statistical and systematic uncertainty (30\%) from the theory. For those cases when either the total or $\gamma$-ray widths are determined, the associated uncertainties on those widths are included. For states that are not observed in the present experiment, an upper limit was assumed, and the dimensionless reduced proton-width was sampled from a Porter-Thomas distribution characterized by a mean value of $\langle\theta_p^2\rangle=0.0045$ with a factor 3 uncertainty \cite{Pogrebnyak2013}. This dimensionless reduced proton-width is related to the proton width through the definition :  $\Gamma_p = \theta^2\times\Gamma_W$ where $\Gamma_W=2 \hbar^2/(\mu s^2) \times P_\ell(E_r,s)$ is the Wigner limit.

\subsection{Inputs}
The resonance parameters used for the calculation of the \spg\ reaction rates are summarized in Table~\ref{table:resonances} together with available information from the literature. The resonance strengths from the present work are calculated under the assumption that the proton widths are much smaller than the $\gamma$-ray widths leading to $\omega\gamma=0.5(2J+1)\Gamma_p$; this approximation holds true up to \Ercm{500}. Below this energy the \Ercm{18.6} and \Ercm{145.8} resonances have not been observed in the present work and an upper limit for their strength is estimated, as explained in the previous subsection. Resonances at \Ercm{275.5} and \Ercm{528.5} were discarded in the present calculations since no additional evidence for the existence of the associated \ph\ levels have been found apart from their single observation~\cite{Teterin1974} and ~\cite{Moss1969} respectively. The strength of the resonances at \Ercm{423.0} and \Ercm{484.6} was measured directly in a recent work~\cite{Dermigny2020}, and a comparison with the respective strengths derived from the present indirect measurement shows a very good agreement within 50\% uncertainty. However, for the reaction rate calculations we preferentially use the resonance strengths obtained from direct measurement~\cite{Dermigny2020}.

\begin{table*}[!hbtp]
    \caption{\label{table:resonances}
    Properties of resonances above the  $^{30}$Si+$p$ threshold from the present work and comparison with the literature.}
    \begin{ruledtabular}
	\begin{tabular}{llcccccccr}
	    \multirow{2}{*}{$E_{x}$ (keV)} & \multirow{2}{*}{$E_r^{c.m.}$ (keV)} & \multicolumn{4}{c}{Present work} & \multicolumn{4}{c}{Literature} \\ 
		\cline{3-6} \cline{7-10}
		 &  &  $\ell$ & (2$J$+1)$C^2S$ &  $\Gamma_{p}$ (eV) & $\omega\gamma$ (eV)\footnotemark[1] &  $J
		^{\pi}$ &  $\ell$ &  $\omega\gamma$ (eV) & Ref. \\
		\hline
		
		7313.7 \emph{16} & 18.6 \emph{16} &   &      & $1.45\times10^{-35}$ \footnotemark[2] &	$1.45\times10^{-35}$ 	& (1/2, 3/2)$^+$ & 0, 2 &  $\leq $6.50 $\times$10$^{-37} $ & \cite{ENDT1990}\\
		
	    7316.2 \emph{ 9} & 19.6 \emph{ 9} & 3 & 0.0075 & $2.94\times10^{-39}$ & $1.18 \times 10^{-38}$ & (5/2, 7/2)$^-$ & 3 &  $\approx$ $8.60 \times 10^{-40}$ & \cite{ENDT1990}\\
	    
	    7347.0 \emph{12} & 50.5 \emph{12} & 1 & 0.0007 & $5.20\times10^{-21}$ & $1.04\times10^{-20}$ & (3/2, 5/2)$^-$ & 1, 3 &  $\leq 5.04 \times 10^{-17}$ & \cite{ENDT1990} \\
	    7356 \emph{16} & 59.4 \emph{16} &  & & & & & & & \\
	    7442.3  \emph{3}  &  145.8 \emph{3} &             &  & $4.33\times10^{-17}$\footnotemark[2] & $2.60\times10^{-16}$ & $11/2^+$  & 6  & $\leq 1.24 \times 10^{-15}$ & \cite{ENDT1990}\\
	    
	    7445.7 \emph{27} \footnotemark[3] & 149.2 \emph{29} & 2 & $0.0007$ & $1.16\times 10^{-11}$  & $2.33\times 10^{-11}$ & (3/2$^+$, 5/2, 7/2, 9/2$^+$) & 2, 3, 4 &   $\leq7.60\times10^{-8}$ & \cite{ENDT1990}\\
	    
	    7470.5 \emph{23} &  174.0 \emph{23}  & 3 & 0.001 & $1.59\times 10^{-12}$ & $6.38\times10^{-12}$  & (7/2, 9/2)$^-$ & 3, 5 &  $\leq1.27\times10^{-10}$   & \cite{ENDT1990}\\
	    
	    7572             & 275.5                &   &        &                &   &       &              &       \\
	     
	    7691.1 \emph{10} & 394.6 \emph{10} & 3 & 0.006 & $1.47\times10^{-6}$ & $4.40\times10^{-6}$    & & &      & \cite{ENDT1990} \\
	    
	    7719.5 \emph{ 8} & 423.0 \emph{ 8} & 3 & 0.044  & $2.48\times10^{-5}$ & $7.42\times10^{-5}$ & (3/2,5/2)$^-$  &       &    ($1.14\pm0.25$) $\times 10^{-4}$  & \cite{Dermigny2020} \\
	    
	    7737.3 \emph{ 7} & 440.8  \emph{ 7} & 3 & 0.114 & $9.79\times10^{-5}$ & $3.92\times10^{-4}$ & (5/2, 7/2)$^-$ & 3 &   $\approx3.72\times10^{-4}$  & \cite{ENDT1990}  \\
	    
	    7781.1 \emph{ 8} & 484.6 \emph{ 8} & 1 & 0.015  &  0.061 & 0.123 & 3/2$^-$ &   &  $0.188\pm0.014$ & \cite{Dermigny2020} \\
	    
	    7825   \emph{ 9}&  528.5 \emph{ 9}&          &           &   &   &       &  &      & \\
	    
	    7851.4 \emph{ 8} & 554.9 \emph{ 8} &   1 & 0.009 & 0.244& 0.181\footnotemark[4] &   &       &  & \\
	    
	    7863.4 \emph{14} & 566.9 \emph{16} & 3 & 0.004 & $5.55\times10^{-5}$ & $1.67\times10^{-4}$  &       &  &  & \\
	    7897.8 \emph{ 7} & 601.3 \emph{ 7} & 1 & 0.115  & 6.49 &  & 1/2$^-$ &  &   $1.95\pm0.10$ & \cite{PAINE1979} \\
	\end{tabular}
	\end{ruledtabular}
	    \footnotetext[1]{Assuming $\Gamma_p\ll\Gamma_\gamma$, so that $\omega\gamma=0.5(2J+1)\Gamma_p$.}
	    \footnotetext[2]{A dimensionless reduced width $\langle\theta^2_p\rangle=0.0045$ is assumed.}
	    \footnotetext[3]{7441.4~keV in literature}
	    \footnotetext[4]{The complete resonance strength formula has been used (see text).}
\end{table*}

For energies above \Ercm{500} the $\Gamma_p\ll\Gamma_\gamma$ assumption may fail and the resonance strength should be computed using Eq.~\ref{eq:omega_gamma}. This is the case of the \Ercm{554.9} resonance for which the proton width ($\Gamma_p=0.24$~eV) and the $\gamma$-ray width ($\Gamma_\gamma=0.7$~eV) are similar. For the resonance at \Ercm{566.9}, no information on the total and $\gamma$-ray width is available, thus the $\Gamma_p\ll\Gamma_\gamma$ approximation has been used to calculate its strength. Note that in any case this assumption will have no impact on the contribution of this resonance to the reaction rate because of the high centrifugal barrier for the incoming proton ($\ell_p=3$) to overcome. Concerning the \Ercm{601.3} resonance: its total width is given by the proton width so the resonance strength is proportional to the $\gamma$-ray width, which is unknown. In this case we use the existing resonance strength obtained from direct measurements.

For resonances above \Ercm{600}, we follow the prescription given in Ref.~\cite{Dermigny2020}. The values of the strengths are taken from the compilation of Ref.~\cite{ENDT1990} and renormalized to the resonance strength of the resonance at \Ercm{602} measured in Ref.~\cite{PAINE1979}.

The direct capture component of the \spg\ reaction rate has also been included. In that case the astrophysical $S$-factor is expanded as $S(E) = S(0) + S'(0)E + S''(0)E^2$ where the energy-independent coefficients are taken from Ref.~\cite{Dermigny2020}.

\subsection{Results and discussion}
\begin{figure}[htbp]
\centering
\includegraphics[width=\columnwidth]{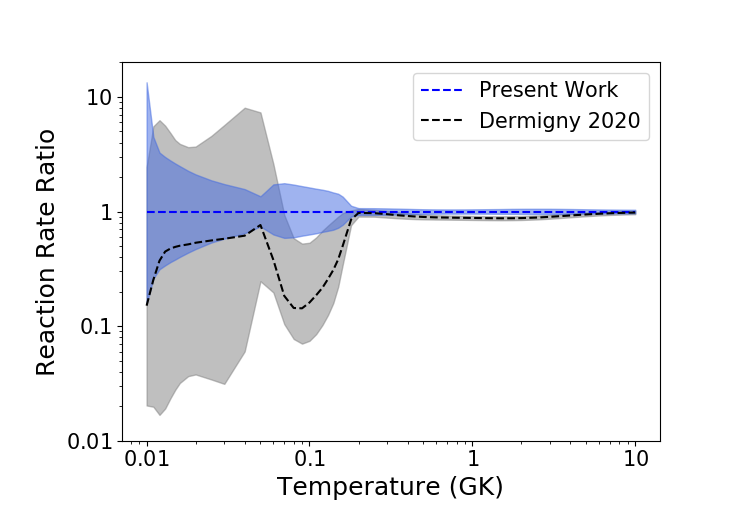}
\caption{(Color online) Reaction rates from the present work (blue) and the evaluation of Dermigny \textit{et al.}~\cite{Dermigny2020}~(gray) normalized to the present recommended rate. The shaded areas represent 68\% coverage probabilities. The \Ercm{149} keV resonance was assumed to have $\ell_p = 2$ in this calculation. The uncertainty of this $\ell$ assignment leads to an additional uncertainty around 0.1 GK (see Fig. 7 and text for details).}\label{fig:rate}
\end{figure}

The results of the Monte Carlo calculations for the \spg~reaction rates are presented in Tables~\ref{tab:rate} and~\ref{tab:rate_bis}. The resonance parameters were sampled 20000 times allowing for the determination of reaction rates with statistical meaningful uncertainties. The recommended value at each temperature is the median value (0.5 quantile) of the cumulative reaction rate distribution, while the low and the high rates represent the 0.16 and 0.84 quantiles, respectively. Fig.~\ref{fig:rate} shows (in blue) these reaction rates normalized to the recommended one as a function of the temperature and the colored area represents a coverage probability of 68\%. The reaction rates from the most recent \spg\ evaluation~\cite{Dermigny2020} are displayed in grey color normalized to the present recommended value. 

\begin{table}
    \caption{Thermonuclear reaction rates for the \spg\ reaction as a function of the temperature in units of cm$^3$~mol$^{-1}$~s$^{-1}$. The \Ercm{149} resonance was assumed to have $\ell_p = 2$ in this calculation.}
\begin{tabular*}{\linewidth}{@{\extracolsep{\fill} }lccc}
\hline \hline
T (GK) & Low & Recommended & High \\
\hline 
0.010&$7.47\times 10^{-38}$&$4.87\times 10^{-37}$&$6.23\times 10^{-36}$\\
0.011&$5.34\times 10^{-36}$&$2.07\times 10^{-35}$&$9.20\times 10^{-35}$\\
0.012&$2.86\times 10^{-34}$&$9.32\times 10^{-34}$&$3.06\times 10^{-33}$\\
0.013&$1.00\times 10^{-32}$&$3.03\times 10^{-32}$&$9.05\times 10^{-32}$\\
0.014&$2.30\times 10^{-31}$&$6.48\times 10^{-31}$&$1.82\times 10^{-30}$\\
0.015&$3.53\times 10^{-30}$&$9.41\times 10^{-30}$&$2.50\times 10^{-29}$\\
0.016&$3.87\times 10^{-29}$&$9.77\times 10^{-29}$&$2.47\times 10^{-28}$\\
0.018&$2.08\times 10^{-27}$&$4.79\times 10^{-27}$&$1.11\times 10^{-26}$\\
0.020&$4.93\times 10^{-26}$&$1.06\times 10^{-25}$&$2.28\times 10^{-25}$\\
0.025&$1.41\times 10^{-23}$&$2.66\times 10^{-23}$&$5.05\times 10^{-23}$\\
0.030&$5.78\times 10^{-22}$&$1.01\times 10^{-21}$&$1.76\times 10^{-21}$\\
0.040&$5.72\times 10^{-20}$&$8.92\times 10^{-20}$&$1.41\times 10^{-19}$\\
0.050&$1.45\times 10^{-18}$&$1.97\times 10^{-18}$&$2.70\times 10^{-18}$\\
0.060&$6.00\times 10^{-17}$&$9.48\times 10^{-17}$&$1.64\times 10^{-16}$\\
0.070&$2.22\times 10^{-15}$&$3.77\times 10^{-15}$&$6.73\times 10^{-15}$\\
0.080&$3.92\times 10^{-14}$&$6.55\times 10^{-14}$&$1.13\times 10^{-13}$\\
0.090&$3.70\times 10^{-13}$&$6.04\times 10^{-13}$&$1.01\times 10^{-12}$\\
0.100&$2.24\times 10^{-12}$&$3.56\times 10^{-12}$&$5.82\times 10^{-12}$\\
0.110&$9.79\times 10^{-12}$&$1.51\times 10^{-11}$&$2.41\times 10^{-11}$\\
0.120&$3.34\times 10^{-11}$&$5.06\times 10^{-11}$&$7.85\times 10^{-11}$\\
0.130&$9.50\times 10^{-11}$&$1.41\times 10^{-10}$&$2.13\times 10^{-10}$\\
0.140&$2.38\times 10^{-10}$&$3.43\times 10^{-10}$&$5.08\times 10^{-10}$\\
0.150&$5.52\times 10^{-10}$&$7.70\times 10^{-10}$&$1.11\times 10^{-09}$\\
0.160&$1.33\times 10^{-09}$&$1.75\times 10^{-09}$&$2.36\times 10^{-09}$\\
0.180&$1.42\times 10^{-08}$&$1.57\times 10^{-08}$&$1.76\times 10^{-08}$\\
0.200&$2.05\times 10^{-07}$&$2.19\times 10^{-07}$&$2.35\times 10^{-07}$\\
0.250&$3.96\times 10^{-05}$&$4.24\times 10^{-05}$&$4.54\times 10^{-05}$\\
0.300&$1.40\times 10^{-03}$&$1.49\times 10^{-03}$&$1.59\times 10^{-03}$\\
0.350&$1.81\times 10^{-02}$&$1.92\times 10^{-02}$&$2.04\times 10^{-02}$\\
0.400&$1.26\times 10^{-01}$&$1.33\times 10^{-01}$&$1.40\times 10^{-01}$\\
0.450&$5.75\times 10^{-01}$&$6.05\times 10^{-01}$&$6.35\times 10^{-01}$\\
0.500&$1.95\times 10^{+00}$&$2.04\times 10^{+00}$&$2.14\times 10^{+00}$\\
0.600&$1.22\times 10^{+01}$&$1.28\times 10^{+01}$&$1.33\times 10^{+01}$\\
0.700&$4.52\times 10^{+01}$&$4.72\times 10^{+01}$&$4.93\times 10^{+01}$\\
0.800&$1.20\times 10^{+02}$&$1.25\times 10^{+02}$&$1.31\times 10^{+02}$\\
0.900&$2.55\times 10^{+02}$&$2.66\times 10^{+02}$&$2.78\times 10^{+02}$\\
1.000&$4.62\times 10^{+02}$&$4.82\times 10^{+02}$&$5.04\times 10^{+02}$\\
1.250&$1.32\times 10^{+03}$&$1.38\times 10^{+03}$&$1.45\times 10^{+03}$\\
1.500&$2.62\times 10^{+03}$&$2.75\times 10^{+03}$&$2.89\times 10^{+03}$\\
1.750&$4.23\times 10^{+03}$&$4.43\times 10^{+03}$&$4.68\times 10^{+03}$\\
2.000&$6.02\times 10^{+03}$&$6.32\times 10^{+03}$&$6.68\times 10^{+03}$\\
2.500&$9.86\times 10^{+03}$&$1.04\times 10^{+04}$&$1.10\times 10^{+04}$\\
3.000&$1.38\times 10^{+04}$&$1.45\times 10^{+04}$&$1.53\times 10^{+04}$\\
3.500&$1.77\times 10^{+04}$&$1.86\times 10^{+04}$&$1.96\times 10^{+04}$\\
4.000&$2.16\times 10^{+04}$&$2.26\times 10^{+04}$&$2.38\times 10^{+04}$\\
5.000&$2.90\times 10^{+04}$&$3.02\times 10^{+04}$&$3.16\times 10^{+04}$\\
6.000&$3.57\times 10^{+04}$&$3.71\times 10^{+04}$&$3.87\times 10^{+04}$\\
7.000&$4.14\times 10^{+04}$&$4.29\times 10^{+04}$&$4.46\times 10^{+04}$\\
8.000&$4.60\times 10^{+04}$&$4.77\times 10^{+04}$&$4.95\times 10^{+04}$\\
9.000&$4.96\times 10^{+04}$&$5.13\times 10^{+04}$&$5.32\times 10^{+04}$\\
10.000&$5.22\times 10^{+04}$&$5.41\times 10^{+04}$&$5.60\times 10^{+04}$\\
\hline \hline
\end{tabular*}
	\label{tab:rate}
\end{table}

\begin{table}
    \caption{Thermonuclear reaction rates for the \spg\ reaction as a function of the temperature in units of cm$^3$~mol$^{-1}$~s$^{-1}$. The \Ercm{149} resonance was assumed to have $\ell_p = 3$ in this calculation and the temperature range is restricted to where the rates are different with respect to those displayed in Table~\ref{tab:rate}.}
\begin{tabular*}{\linewidth}{@{\extracolsep{\fill} }lccc}
\hline \hline
T (GK) & Low & Recommended & High \\
\hline 
0.050&$1.17\times 10^{-18}$&$1.59\times 10^{-18}$&$2.22\times 10^{-18}$\\
0.060&$1.91\times 10^{-17}$&$2.55\times 10^{-17}$&$3.44\times 10^{-17}$\\
0.070&$3.15\times 10^{-16}$&$4.36\times 10^{-16}$&$6.08\times 10^{-16}$\\
0.080&$4.04\times 10^{-15}$&$5.54\times 10^{-15}$&$7.69\times 10^{-15}$\\
0.090&$3.58\times 10^{-14}$&$4.81\times 10^{-14}$&$6.60\times 10^{-14}$\\
0.100&$2.29\times 10^{-13}$&$3.07\times 10^{-13}$&$4.16\times 10^{-13}$\\
0.110&$1.13\times 10^{-12}$&$1.50\times 10^{-12}$&$2.02\times 10^{-12}$\\
0.120&$4.49\times 10^{-12}$&$6.02\times 10^{-12}$&$8.14\times 10^{-12}$\\
0.130&$1.53\times 10^{-11}$&$2.04\times 10^{-11}$&$2.80\times 10^{-11}$\\
0.140&$4.70\times 10^{-11}$&$6.31\times 10^{-11}$&$8.54\times 10^{-11}$\\
0.150&$1.47\times 10^{-10}$&$1.90\times 10^{-10}$&$2.53\times 10^{-10}$\\
0.160&$5.48\times 10^{-10}$&$6.58\times 10^{-10}$&$8.16\times 10^{-10}$\\
0.180&$1.16\times 10^{-08}$&$1.26\times 10^{-08}$&$1.36\times 10^{-08}$\\
0.200&$1.98\times 10^{-07}$&$2.12\times 10^{-07}$&$2.27\times 10^{-07}$\\
\hline \hline
\end{tabular*}
	\label{tab:rate_bis}
\end{table}

The recommended reaction rate calculated in the present work is two times higher than the rate of Dermigny {\it et al.}~\cite{Dermigny2020} for $T \leq 50~\mathrm{MK}$. This is due to the experimental determination of the spectroscopic factor of the resonance at \Ercm{50.5}, which dominates the rate at these temperatures. In the work of Dermigny {\it et al.}, the \Ercm{50.5} resonance was considered as an upper limit with the reduced proton width sampled from a Porter-Thomas distribution with a mean reduced width of $0.0003$ and a factor 3 associated uncertainty. 
The determination of this spectroscopic factor and thus proton width leads to an order of magnitude reduction in the uncertainty in the reaction rate since the upper limits sampled from a Porter Thomas distribution span a much larger range of values than for a log-normal distribution associated to a strength measurement. In the temperature range $50~\mathrm{MK} \leq T \leq 200~\mathrm{MK}$, our rates are up to seven times higher, because of the observation of the \Ercm{149} and \Ercm{174} resonances in the present work. These resonances were treated in Dermigny {\it et al.} as upper limits, as for the \Ercm{50.5} resonance. In addition, several different spin assignments were considered in the case of the \Ercm{149} resonance~\cite{Dermigny2020} (see later for discussion). For temperatures above 200~MK, the present rate is identical to that of Dermigny~{\it et al.}~\cite{Dermigny2020} since the same resonance strengths have been used for the higher energy resonances.

\begin{figure}[htbp]
    \centering
    \includegraphics[width=\columnwidth]{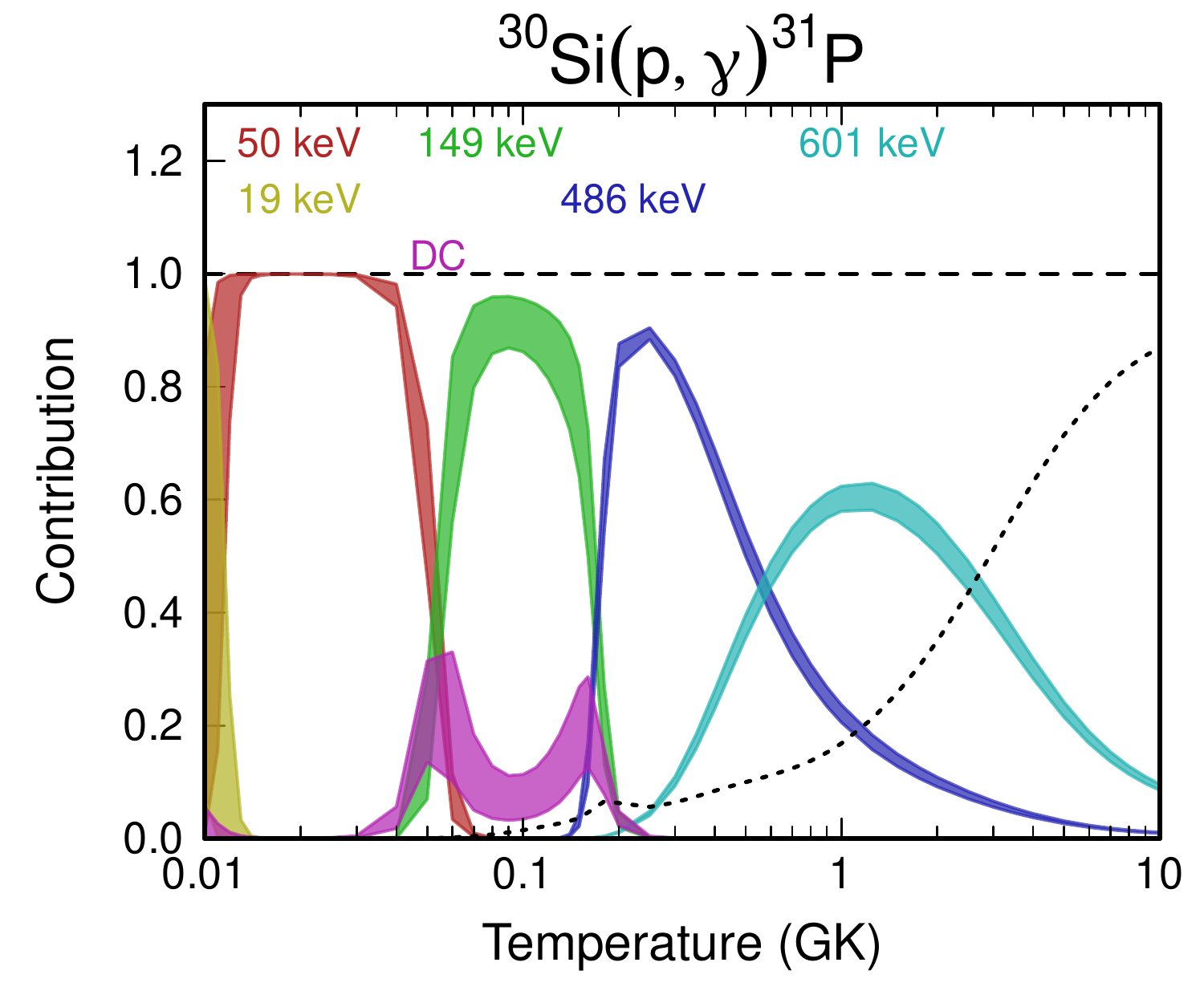}
    \caption{(Color online) Fractional contribution of \spg\ resonances, along with direct capture contribution (labeled "DC") to the total reaction rate. The thickness of each band represents the uncertainty of the contribution. The dotted black line represents the contribution to the total rate of resonances with energies above \Ercm{648}. The \Ercm{149} resonance was assumed to have $\ell_p = 2$ in this calculation (see text for details).
    \label{fig:frac_contribution}}
\end{figure}

The fractional contribution of each resonance to the reaction rate at each temperature, along with the direct capture contribution, is shown in Fig.~\ref{fig:frac_contribution}. Noteworthy is that the rate of the \spg\ reaction is mainly dominated by four observed resonances. For the two higher-energy resonances (\Ercm{485} and \Ercm{601}) the strengths have been measured directly. The strengths of the \Ercm{50} and \Ercm{149} resonances have been obtained indirectly in the present study. In the temperature range of $120-200$~MK relevant for hydrogen burning in the polluting stellar site in globular clusters, the \spg\ reaction rate is dominated by the \Ercm{149} resonance.

\begin{figure}[htbp]
    \centering
    \includegraphics[width=\columnwidth]{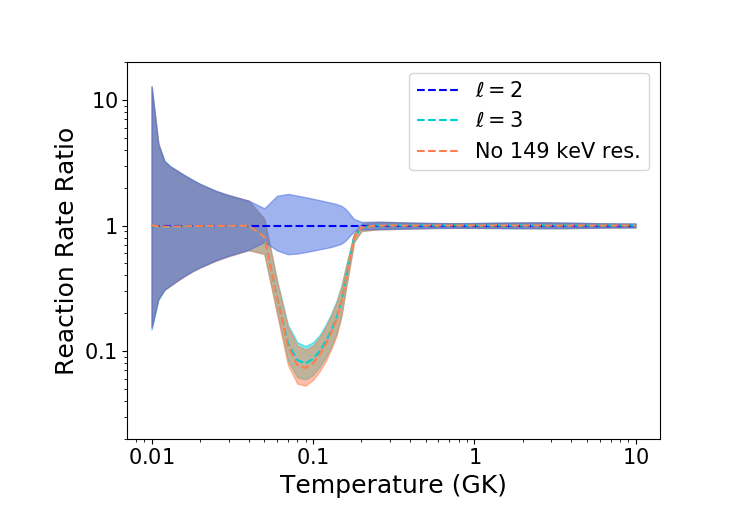}
    \caption{(Color online) Thermonuclear rates of the \spg\ reaction calculated for different transferred orbital angular momenta for the resonance at \Ercm{149}. All rates are normalised to the recommended value. \label{fig:L_comparaison}}
\end{figure}

The analysis of the angular distribution of the \ph\ state at 7446~keV corresponding to the \Ercm{149} resonance does not provide a unique determination of the transferred angular momentum (see Sec.~\ref{sec:149}). The reaction rates presented in Fig.~\ref{fig:rate} assume an $\ell=2$ transfer, however $\ell=3$ could also be possible. The impact of this situation has been explored and results are presented in Fig.~\ref{fig:L_comparaison} where three sets of reaction rates have been calculated assuming $\ell=2$, $\ell=3$ and no contribution at all; the three cases are normalized to the $\ell=2$ recommended reaction rate. A factor of ten difference is observed between the $\ell=2$ and $\ell=3$ cases. In addition, the $\ell=3$ case gives a very similar reaction rate as when the resonance is switched off because the direct capture component starts to be the dominant contribution. Interestingly, the \spg\ rate calculated with $\ell=3$ is very close to Dermigny~\textit{et al.}~\cite{Dermigny2020} rate (see Fig.~\ref{fig:rate}). In their work the possible proton orbital angular momentum of the \Ercm{149} resonance is sampled with equal probability between $\ell=2$, $\ell=3$ and $\ell=4$. Given that the recommended reaction rate is defined as the median of the cumulative distribution function (50$^{th}$ percentile) the Dermigny~\textit{et al.}~\cite{Dermigny2020} recommended rate naturally corresponds to the $\ell=3$ case. These results stress the importance to determine the spin and parity of the \Ercm{149} resonance in order to constrain the \spg\ rate in the temperature range 50~MK~$\leq T\leq$~200~MK which is relevant for understanding the nature of the polluting stars in globular clusters. We chose to present the different rates separately in Table~\ref{tab:rate} for $\ell=2$ and Table~\ref{tab:rate_bis} for $\ell=3$ where the rates are different.

\section{Summary and conclusions}
\label{sec:conclusion}
A high-resolution measurement of the one-proton \shd\ transfer reaction was performed with the tandem and Q3D spectrometer at the Maier-Leibnitz-Laboratorium in Munich. States with excitation energies between 6.8~MeV and 8.1~MeV have been populated. The doublet of states at 7719~keV and 7737~keV could be resolved, and the states at 7347, 7446, 7470, 7691 and 7863~keV were observed for the first time in a one-proton transfer reaction. Angular distributions have been extracted for 25 \ph\ states and proton spectroscopic factors have been obtained using a finite range DWBA analysis, some of them for the first time. Proton widths were subsequently calculated for unbound states and the strengths of \nuc{30}{Si}+$p$ resonances were determined up to \Ercm{600}.

An updated thermonuclear \spg\ reaction rate with its corresponding statistical uncertainty has been obtained using a Monte-Carlo approach. With the first determination of proton widths reported in the present work for the resonances at \Ercm{50} and \Ercm{149} the \spg\ reaction rate is now fully based on measured quantities. In the temperature range of $120-200$~MK relevant for hydrogen burning in the polluting stellar site in globular clusters, the \spg\ reaction rate is dominated by the \Ercm{149} resonance. The spin and parity of this resonance are unknown and this represents the main remaining uncertainty in the reaction rate. Whether the proton capture occurs through a $d$- or $f$-wave for this resonance induces a factor of 10 uncertainty in the reaction rate. However, once its spin and parity are determined the uncertainty on the reaction rate will be reduced to a factor of two. Thus, we strongly encourage new indirect studies aimed at the determination of the spin and parity of the \nuc{30}{Si}+$p$ resonance at \Ercm{149}.

\begin{acknowledgements}
It is a pleasure to be able to thank the beam operators at the Maier-Leibnitz-Laboratorium, Munich for the high-quality beam produced. NdS thanks Federico Portillo Chaves and Kiana Setoodehnia for their assistance in producing high quality enriched targets. The authors also acknowledge the INFN-LNS target laboratory. PA thanks the trustees and staff of the Claude Leon Foundation for support in the form of a Postdoctoral Fellowship. RL is supported by the U.S. Department of Energy, Office of Science, Office of Nuclear Physics, under award number DE-SC0017799 and contract number DE-FG02-97ER41041.
\end{acknowledgements}

\bibliographystyle{apsrev4-1}
\bibliography{si30he3d}

\end{document}